\documentclass[twocolumn,english,prb,amsmath,amssymb,eqsecnum,preprintnumbers]{revtex4}
\usepackage[T1]{fontenc}
\usepackage[latin9]{inputenc}
\usepackage{color}
\usepackage{bm}
\usepackage{amstext}
\usepackage{esint}
\usepackage{soul}
\usepackage{graphicx}
\usepackage{dcolumn}
\usepackage{bigstrut}
\usepackage{rotating} 
\usepackage{yfonts}   

\makeatletter
\@ifundefined{textcolor}{}
{%
 \definecolor{BLACK}{gray}{0}
 \definecolor{WHITE}{gray}{1}
 \definecolor{RED}{rgb}{1,0,0}
 \definecolor{GREEN}{rgb}{0,1,0}
 \definecolor{BLUE}{rgb}{0,0,1}
 \definecolor{CYAN}{cmyk}{1,0,0,0}
 \definecolor{blue}{cmyk}{0,1,0,0}
 \definecolor{YELLOW}{cmyk}{0,0,1,0}
}

\def\vF{v_{\text{F}}}

\def\be{\begin{equation}}
\def\ee{\end{equation}}
\def\bea{\begin{eqnarray}}
\def\eea{\end{eqnarray}}
\def\bse{\begin{subequations}}
\def\ese{\end{subequations}}

\def\bml{\begin{mathletters}}
\def\eml{\end{mathletters}}

\usepackage{dcolumn}
\usepackage{bm}
\usepackage{amsfonts}
\usepackage[normalem]{ulem}
\draft

\makeatother

\usepackage{babel}
\begin{document}

\title{Spin dynamics of antiferromagnets in the presence of a homogeneous magnetization}

\author{T. R. Kirkpatrick$^{1}$ and D. Belitz$^{2,3}$}

\affiliation{ $^{1}$ Institute for Physical Science and Technology, University of Maryland, College Park, MD 20742, USA\\
                  $^{2}$ Department of Physics and Institute of Theoretical Science, University of Oregon, Eugene, OR 97403, USA\\
                  $^{3}$ Materials Science Institute, University of Oregon, Eugene, OR 97403, USA\\
}

\date{\today}
\begin{abstract}
We use general hydrodynamic equations to determine the long-wavelength spin excitations in isotropic antiferromagnets in the
presence of a homogeneous magnetization. The latter may be induced, such as in antiferromagnets in an external
magnetic field, or spontaneous, such as in ferrimagnetic or canted phases that are characterized by the coexistence of
antiferromagnetic and ferromagnetic order. Depending on the physical situation, we
find propagating spin waves that are gapped in some cases and gapless in others, diffusive modes, or relaxational
modes. The excitation spectra turn out to be qualitatively different depending on whether or not the homogeneous
magnetization is a conserved quantity. The results lay the foundation for a description of a variety of quantum phase
transitions, including the transition from a ferromagnetic metal to an antiferromagnetic one, and the spin-flop
transitions that are observed in some antiferromagnets. They also are crucial for incorporating weak-localization and
Altshuler-Aronov effects into the descriptions of quantum phases in both clean and disordered magnetic metals.
\end{abstract}
%
%
\maketitle

\section{Introduction}
\label{sec:I}

Soft or massless excitations are of paramount importance for the description of condensed-matter systems,
since they determine the universal long-wavelength and low-frequency properties of materials that do not
depend on microscopic details. A common cause of soft excitations or modes is the presence of a
spontaneously broken continuous symmetry in an ordered phase, which leads to static Goldstone modes
and related dynamical excitations.\cite{Forster_1975} They generically couple to various observables and
qualitatively change the behavior of both static susceptibilities and equilibrium time-correlation functions;
namely, they can lead to power-law instead of exponential decay for large distances or times, a phenomenon
known as generic scale invariance.\cite{Belitz_Kirkpatrick_Vojta_2005}

In magnets, the excitations due to the long-ranged magnetic order are magnons or spin-waves. In simple
isotropic ferromagnets and antiferromagnets they are well known to be gapless with a quadratic and linear dispersion
relation, respectively, in the long-wavelength limit. This difference is due to the coupling of the antiferromagnetic
order parameter, i.e., the staggered magnetization, to the fluctuating homogeneous magnetization, as a consequence of
which the two problems do not simply map onto one another.\cite{Forster_1975} 

For antiferromagnets in the presence of a nonzero average homogeneous magnetization, which can be due to the 
presence of an external magnetic field or coexisting spontaneous ferromagnetic and antiferromagnetic orders, a 
systematic analysis of the soft modes, or spin excitations in general, in the ordered phase does not exist. This is
rather surprising, given the abundance of antiferromagnetic materials, and the importance of the concept for
many topics of great current interest, including certain classes of quantum phase transitions\cite{Gegenwart_Si_Steglich_2008}
and high-$T_{\rm c}$ as well as iron-based superconductors.\cite{Lee_Nagaosa_Wen_2006, Stewart_2011} Early work on the dynamics of
antiferromagnets focused on the dynamical critical behavior\cite{Freedman_Mazenko_1976, Hohenberg_Halperin_1977}
and therefore left out terms that renormalization-group irrelevant near the classical critical point, and later
approaches that used spin-wave theory or other solid-state oriented approaches were not systematic and
sometimes reached conclusions that are not consistent with basic spin dynamics. 

It is the purpose of the present paper to remedy this situation and give a complete classification of the long-wavelength
spin excitations in antiferromagnets in the presence of a homogeneous magnetization. We use a 
hydrodynamic approach that is extremely general and reliable and has been previously applied 
to helical magnets,\cite{Belitz_Kirkpatrick_Rosch_2006a, Kirkpatrick_Belitz_2006} and to ferromagnets.\cite{Bharadwaj_Belitz_Kirkpatrick_2016}
As we will see, it relies only on the basic equation of motion for a magnetic moment and therefore
is more general than approaches based on specific solid-state-oriented models, such as the Heisenberg model.
For simplicity, we consider only the case of isotropic magnets, as the problem is fairly complex even
in that simple case. If desirable, the symmetry-breaking effects of the spin-orbit interaction can be
built in at a later stage. Our results are very general and depend only on the type of order and on
conservation laws, rather than on the underlying mechanisms that produce the order. For instance,
the spin-wave spectrum is the same irrespective of whether the staggered magnetization and the
homogeneous part of the order parameter are produced by electrons in the same band or electrons
in different bands, and it the same in what are known as ``canted phases'' and ``fan phases''.\cite{Chaikin_Lubensky_1995}
As we will show, the results are qualitatively different depending on whether the homogeneous
magnetization is conserved, or whether that conservation law is violated, e.g., due to the presence
of magnetic impurities.

To conclude these introductory remarks we list some physical problems for which a thorough 
understanding of the spin dynamics is crucial.

(1) In ferromagnets, and in simple antiferromagnets, the spin waves couple to other observables, e.g.,
the longitudinal susceptibility and the dynamical structure factor, which is directly observable via
neutron scattering. This coupling induces nonanalytic wave-number and frequency dependences
that reflect the long-range order in the magnetic phase.\cite{Bharadwaj_Belitz_Kirkpatrick_2016} 
Similar effects are expected for the more complicated antiferromagnets discussed here.

(2) More generally, the effects known as weak-localization phenomena and Altshuler-Aronov effects
in disordered metals,\cite{Altshuler_Aronov_1984, Lee_Ramakrishnan_1985, Belitz_Kirkpatrick_Vojta_2005}
as well as their counterparts in clean metals,\cite{Belitz_Kirkpatrick_Vojta_1997, Belitz_Kirkpatrick_2014} rely on all of the soft modes
in the system and their couplings to various observables. A complete list of soft modes is therefore
crucial for studying these effects, and in magnetic metals this includes the soft collective spin
excitations.

(3) At the phase transition that signals the instability of an ordered phase, the soft modes that
characterize the latter disappear, turn into critical modes, or change into modes characteristic
of a different type of order, depending on the nature of the phase transition. Knowledge of the
soft modes is thus important for describing the transition. In the current context, an interesting
example is the quantum phase transition from a ferromagnetic phase to an antiferromagnetic
one. There are many experimental examples of such transitions,\cite{Brando_et_al_2016a}
but no theoretical description exists. Another example of phase transitions for which information
about spin waves is important are the metamagnetic transitions known as spin-flop transitions
(e.g., from easy-axis to easy-plane) that are commonly observed in antiferromagnets, see
Ref.~\onlinecite{Bogdanov_Zhuravlev_Rossler_2007} and references therein.

(4) If antiferromagnetic spin fluctuations are behind the pairing mechanism for either high-$T_{\rm c}$ 
or iron-based superconductors, 
as has been suggested,\cite{Norman_2007, Stewart_2011} then one would expect their spectra to be reflected in
tunneling data, just as is the case for the phonon spectra in conventional superconductors. 
Moreover, the spin fluctuations responsible for the pairing mechanism would be very sensitive
to an external magnetic field, which is not the case for phonons.  A thorough understanding of spin 
fluctuations, especially in an external field, is therefore very important in this context. We note
that coexistence of antiferromagnetic order and superconductivity has been observed in some
materials, see Ref.~\onlinecite{Uemura_2014} and references therein, which will make spin
excitations in a magnetically ordered state directly relevant. However, even in cases where
there is an antiferromagnetic phase nearby in the phase diagram a thorough understanding
of the antiferromagnetic parent compound is important. 

(5) While magnetic states that have both a antiferromagnetic and a ferromagnetic component have
been known for a long time, materials that display such phases have received much attention
lately, in part because of their potential technological importance, see, e.g., Refs.~\onlinecite{Marti_et_al_2010, White_et_al_2013}.
Their understanding requires information about the spin dynamics of systems in which both
order parameters are nonzero.

\section{Time-dependent Ginzburg-Landau theory}
\label{sec:II}

While the equations of motion for an isotropic Heisenberg antiferromagnet are well known,\cite{Hohenberg_Halperin_1977, Forster_1975,
Chaikin_Lubensky_1995} many versions in the literature
omit terms that are irrelevant for the classical critical behavior, yet contribute to the spin dynamics in the ordered phase.
For completeness, we therefore provide a brief derivation. 

\subsection{Statics}
\label{subsec:II.A}

\subsubsection{Landau free energy}
\label{subsubsec:II.A.1}

Consider a general magnetization field ${\bm M}({\bm x})$ of the form
\bse
\label{eqs:2.1}
\be
{\bm M}({\bm x}) = {\bm m}({\bm x}) + {\bm\nu}({\bm x})\ ,
\label{eq:2.1a}
\ee
where ${\bm m}({\bm x})$ is a slowly varying function, whereas
\be
{\bm\nu}({\bm x}) = {\bm n}({\bm x})\,f({\bm x})
\label{eq:2.1b}
\ee
\ese
with ${\bm n}({\bm x})$ slowly varying and $f({\bm x})$ a rapidly oscillating function with zero spatial mean. We defined coarse-grained
variables
\bse
\label{eqs:2.2}
\bea
\frac{1}{V_{\cal N}} \int_{{\bm y}\in {\cal N}({\bm x})} d{\bm y}\,{\bm M}({\bm y}) \approx {\bm m}({\bm x})\ ,
\label{eq:2.2a}\\
\frac{1}{V_{\cal N}}  \int_{{\bm y}\in {\cal N}({\bm x})} d{\bm y}\,{\bm M}({\bm y})\,f({\bm y}) \approx {\bm n}({\bm x})\,\overline{f^2}\ ,
\label{eq:2.2b}
\eea
\ese
where ${\cal N}({\bm x})$ is a neighborhood of the point ${\bm x}$ whose volume $V_{\cal N}$ is large on the microscopic length scale, 
but small on the macroscopic one, and $\overline{f^2}$ is the spatial average of $f^2({\bm x})$. ${\bm m}$ and ${\bm n}$ are the 
magnetization and the staggered magnetization, respectively, and
\bse
\label{eqs:2.3}
\bea
\frac{1}{V_{\cal N}} \int_{{\bm y}\in {\cal N}({\bm x})} d{\bm y}\,\left(f({\bm y})\right)^{n+1} \approx 0 \ ,
\label{eq:2.3.a}\\
\frac{1}{V_{\cal N}} \int_{{\bm y}\in {\cal N}({\bm x})} d{\bm y}\,\left(f({\bm y})\right)^{2n} \equiv \overline{f^{2n}} > 0
\label{eq:2.3.b}
\eea
for all integer $n$. By rescaling ${\bm n}$ we can choose
\be
\overline{f^2} = 1
\label{eq:2.3.c}
\ee
\ese
without loss of generality, and we will adopt this choice from now on. 

Now consider a Landau free energy that is a functional of ${\bm M}$. Multiplying out powers of ${\bm M}$ yields all possible scalar terms that can
be constructed from the two vectors ${\bm m}$ and ${\bm n}$. However, all terms that are odd in ${\bm n}$ are multiplied by odd powers
of $f$ and thus vanish upon coarse graining. Up to quartic terms in ${\bm m}$, ${\bm n}$, and gradients (for comments on higher-order
terms see Sec.~\ref{subsubsec:IV.B.1}) we thus obtain a free-energy
functional 
\bea
F &=& \int d{\bm x}\,\left[\frac{r}{2}\,{\bm n}^2 + \frac{a}{2}\left(\nabla{\bm n}\right)^2 + \frac{u}{4}\left({\bm n}^2\right)^2  - {\textswab{h}}\cdot{\bm n}       
         \right.
\nonumber\\
&& \hskip 30pt + \frac{t}{2}\,{\bm m}^2 + \frac{c}{2}\left(\nabla{\bm m}\right)^2 + \frac{v}{4}\left({\bm m}^2\right)^2  -{\bm h}\cdot{\bm m}
\nonumber\\
&& \hskip 30pt + \frac{1}{2}\,w_1\,{\bm n}^2{\bm m}^2 + \frac{1}{2}\,w_2\, ({\bm n}\cdot{\bm m})^2 \Bigr]\ .\qquad
\label{eq:2.4}
\eea
Here ${\bm h}$ and  ${\textswab{h}}$ are a homogeneous and a staggered magnetic field, respectively, and 
$r$, $a$, $u$, $t$, etc. are Landau coefficients. The coefficients of terms that result from different powers of 
${\bm M}$ are different 
even within a bare theory. Moreover, all of the coefficients will behave differently under renormalization, and 
therefore all Landau coefficients in Eq.~(\ref{eq:2.4}) should be considered independent. In particular, one can
have $r<0$, $t>0$. This allows for spontaneous antiferromagnetic order, i.e., a nonzero staggered magnetization 
for ${\textswab{h}}=0$, with a homogeneous magnetization that vanishes as ${\bm h}\to 0$.
We will also consider the case of coexisting spontaneous order for both the staggered and the homogeneous magnetization.

\subsubsection{Equations of state}
\label{subsubsec:II.A.2}

We now consider the mean-field equations of state, which are given by
\bse
\label{eqs:2.5}
\be
(\delta F/\delta {\bm n})_{{\bm n}_0} = (\delta F/\delta{\bm m})_{{\bm m}_0} = 0\ .
\label{eq:2.5a}
\ee
For ${\textswab{h}} = 0$ we have explicitly
\bea
0 &=& r\,{\bm n_0} + u ({\bm n_0})^2 {\bm n}_0 + w_1 ({\bm m}_0)^2 {\bm n}_0 + w_2({\bm n}_0\cdot{\bm m}_0)\,{\bm m}_0\ ,
\nonumber\\
\label{eq:2.5b}\\
{\bm h} &=&t {\bm m}_0 + v ({\bm m}_0)^2{\bm m}_0 + w_1 ({\bm n}_0)^2{\bm m}_0 
\nonumber\\
&& \hskip 95pt + w_2({\bm n}_0\cdot{\bm m}_0)\,{\bm n}_0\ .
\nonumber\\
\label{eq:2.5c}
\eea
\ese

\paragraph{a. AFM order only}
\label{par:II.A.1.a}

Let us first consider parameter values such that ${\bm n}_0 \neq 0$ and ${\bm m}_0({\bm h}\to 0) \to 0$.
For ${\bm h}=0$ we have the simple AFM solution ${\bm m}_0 = 0$, ${\bm n}_0 = (0,0,n_0)$ with $n_0 = \sqrt{-r/u}$.
For ${\bm h}\neq 0$ we need to distinguish between two cases:
\medskip\par\noindent
{\it Case 1:} $w_2 > 0$
\par
In this case ${\bm n}_0\cdot{\bm m}_0 = 0$, and ${\bm m}_0 = m_0 {\hat{\bm h}}$, with ${\hat{\bm h}}$ the unit vector in
the direction of ${\bm h}$. Choosing ${\bm n}_0 = (0,0,n_0)$, ${\bm m}_0 = (m_0,0,0)$, ${\bm h} = (h,0,0)$, $n_0$ and
$m_0$ are the solutions of the equations of state
\bse
\label{eqs:2.6}
\bea
(t + w_1 n_0^2) m_0 + v m_0^3 = h\ ,
\label{eq:2.2'''a}\\
n_0^2 = -(r + w_1 m_0^2)/u\ ,
\label{eq:2.6b}
\eea
\ese
which requires $r < -w_1 m_0^2$. A Gaussian stability analysis (see Sec.~\ref{subsubsec:II.A.3} below) shows that the condition for this state
to minimize the free energy is
\be
w_1^2 < u\,v + u\,h/2m_0^3\ .
\label{eq:2.7}
\ee
We will refer to this case as the transverse-field case. Note that is the field whose direction is chosen in an experiment,
and $\bm n$ adjusts such that $\bm n$ and $\bm h$ are perpendicular.

\medskip\par\noindent
{\it Case 2:} $w_2 < 0$
\par
In this case ${\bm n}_0$, ${\bm m}_0$, and ${\bm h}$ are all collinear and the equations of state read
\bse
\label{eqs:2.8}
\bea
\left(t + (w_1+w_2) n_0^2\right) m_0 + v m_0^3 = h\ ,
\label{eq:2.8a}\\
n_0^2 = -\left(r + (w_1+w_2) m_0^2\right)/u\ ,
\label{eq:2.8b}
\eea
\ese
which requires $r < -(w_1+w_2) m_0^2$. The stability requirement in this case is
\be
(w_1 + w_2)^2 < u\,v\ .
\label{eq:2.9}
\ee
We will refer to this case as the longitudinal-field case. As in the previous case, ${\bm n}$ will adjust, in this case
such that it is collinear with ${\bm h}$.

\paragraph{b. Coexisting AFM and homogeneous order}
\label{par:II.A.1.b}

Now consider the case ${\bm h}=0$, and parameter values such that both the staggered magnetization and the
homogeneous magnetization have nonzero expectation values, $n_0 \neq 0$ and $m_0 \neq 0$.\cite{FM_footnote}
We need to distinguish again between two cases:
\medskip\par\noindent
{\it Case 1:} $w_2 > 0$,
\par
In this case ${\bm n}_0 \perp {\bm m}_0$. The equations of state are given by Eqs.~(\ref{eqs:2.6}) with $h=0$,
which leads to
\bse
\label{eqs:2.10}
\bea
n_0^2 = -(rv - w_1 t)/(uv - w_1^2)\ ,
\label{eq:2.10a}\\
m_0^2 = -(tu - w_1 r)/(uv - w_1^2)\ .
\label{eq:2.10b}
\eea
\ese
Coexisting orders thus require $rv-w_1t<0$ and $tu-w_1r<0$. The stability criterion is given by Eq.~(\ref{eq:2.7})
with $h=0$, i.e.,
\be
w_1^2 < u\,v\ .
\label{eq:2.11}
\ee
We will refer to this case as the orhogonal-order-parameters case.

\medskip\par\noindent
{\it Case 2:} $w_2 < 0$,
\par
In this case ${\bm n}_0$ is parallel to ${\bm m}_0$. The equations of state are given by Eqs.~(\ref{eqs:2.8})
with $h=0$, which leads to
\bse
\label{eqs:2.12}
\bea
n_0^2 &=& -\left(rv - (w_1+w_2) t\right)/\left(uv - (w_1+w_2)^2\right)\ ,
\nonumber\\
\label{eq:2.12a}\\
m_0^2 &=& -\left(tu - (w_1+w_2) r\right)/\left(uv -( w_1+w_2)^2\right)\ ,
\nonumber\\
\label{eq:2.12b}
\eea
\ese
and the stability criterion is given by Eq.~(\ref{eq:2.9}). We will refer to this case as the collinear-order-parameters case.

\subsubsection{Static susceptibilities; Goldstone modes}
\label{subsubsec:II.A.3}

Also of interest are the static susceptibilities, which are obtained by expanding the free energy to quadratic order in the
Gaussian fluctuations about the solutions of the equations of state. We are in particular interested in the presence of
Goldstone modes, which manifest themselves as susceptibilities that diverge as $k\to 0$. The same calculation yields
the stability criteria listed in Sec.~\ref{subsubsec:II.A.2} above; they are given by the requirement that all of the static
susceptibilities must be positive.

To proceed, we parameterize the fields ${\bm n}$ and ${\bm m}$ as follows:
\be
{\bm n} = (-\theta_2,\theta_1,n_0 + \theta_3)\quad,\quad {\bm m} = {\bm m}_0 + (\pi_1,\pi_2,\pi_3)\ ,
\label{eq:2.13}
\ee
and expand the free energy, Eq.~(\ref{eq:2.4}), to Gaussian order in the small fluctuations $\theta_{1,2,3}$ and $\pi_{1,2,3}$.
This yields a $6\times 6$ eigenvalue problem. Stability requires that all eigenvalues are positive, and eigenvalues that
vanish as $k\to 0$ indicate the existence of Goldstone modes.
\medskip

\paragraph{a. AFM order only}
\label{par:II.A.3.a}
$ $
\medskip\par\noindent
{\it Case 1:} $w_2 > 0$ (transverse-field case)
\par
We choose ${\bm m}_0 = (m_0,0,0)$, and ${\bm h} = (h,0,0)$. The $6\times 6$ problem 
then decomposes into two $2\times 2$ problems (for the pairs $(\theta_2,\pi_3)$ and
$(\theta_3,\pi_1)$, respectively, and two single-variable problems for $\theta_1$ and $\pi_2$, respectively. 
Using the equations of state, Eqs.~(\ref{eqs:2.6}), we find for the Gaussian-fluctuation contribution to the free energy
\begin{widetext}
\bea
\delta F^{(2)} &=& \frac{1}{2V}\sum_{\bm k} \theta_1({\bm k})\,a k^2\,\theta_1(-{\bm k}) + \frac{1}{2V}\sum_{\bm k}  \pi_2({\bm k})\left(h/m_0+ck^2\right)\pi_2(-{\bm k})
\nonumber\\
&& + \frac{1}{2V}\sum_{\bm k}\left(\theta_2({\bm k}),\pi_3({\bm k})\right)\begin{pmatrix} w_2 m_0^2 + a k^2  &  -w_2 n_0 m_0  \\
                                                                                                                                         -w_2 n_0 m_0           &  h/m_0 + w_2 n_0^2 + c k^2 \\
                                                                                                                                         \end{pmatrix}
                                                                                                                 \begin{pmatrix} \theta_2(-{\bm k})  \\
                                                                                                                                          \pi_3(-{\bm k}) \\
                                                                                                                                        \end{pmatrix}
\nonumber\\
 && + \frac{1}{2V}\sum_{\bm k}\left(\theta_3({\bm k}),\pi_1({\bm k})\right)\begin{pmatrix} u n_0^2 + a k^2  &  w_1 n_0 m_0  \\
                                                                                                                                         w_1 n_0 m_0           &  h/m_0 + v m_0^2 + c k^2 \\
                                                                                                                                         \end{pmatrix}
                                                                                                                 \begin{pmatrix} \theta_3(-{\bm k})  \\
                                                                                                                                          \pi_1(-{\bm k}) \\
                                                                                                                                        \end{pmatrix}     \ .                                                                                                                                  
\label{eq:2.14}  
\eea    
\end{widetext}                                                                                                                                      
All eigenvalues are positive provided Eq.~(\ref{eq:2.7}) holds, and the one related to $\theta_1$ vanishes
as $k\to 0$. We thus have one Goldstone mode, 
\bse
\label{eqs:2.15}
\be
g({\bm k}) = \theta_1({\bm k})\ ,
\label{eq:2.15.a}
\ee
whose susceptibility is soft, namely
\be
\chi_{g}(k) = \langle g({\bm k})\,g({\bm k})^*\rangle = 1/a\,k^2\ .
\label{eq:2.15b}
\ee
\ese
Physically, the field
polarizes the homogeneous magnetization ${\bm m}$, and the $w_2$ term forces the staggered magnetization
${\bm n}$ to be perpendicular to ${\bm m}$, but ${\bm n}$ is still free to rotate about the field direction, so one of
the two transverse ${\bm n}$ fluctuations do not cost any energy in the long-wavelength limit.

\medskip\par\noindent
{\it Case 2:} $w_2 < 0$ (longtitudinal-field case)
\par
In this case ${\bm n}_0$, ${\bm m}_0$, and ${\bm h}$ are all collinear, so ${\bm m}_0 = (0,0,m_0)$ and
${\bm h} = (0,0,h)$. The $6\times 6$ problem decomposes into three $2\times 2$ problems for the pairs
$(\theta_1,\pi_2)$, $(\theta_2,\pi_1)$, and $(\theta_3,\pi_3)$, respectively. Using Eqs.~(\ref{eqs:2.8}) we
find for the Gaussian-fluctuation contribution to the free energy
\begin{widetext}
\bea
\delta F^{(2)} &=& \frac{1}{2V}\sum_{\bm k} \left(\theta_1({\bm k}),\pi_2({\bm k})\right)\begin{pmatrix} -w_2 m_0^2 + a k^2  &  w_2 n_0 m_0  \\
                                                                                                                                         w_2 n_0 m_0           &  h/m_0 - w_2 n_0^2 + c k^2 \\
                                                                                                                                         \end{pmatrix}
                                                                                                                 \begin{pmatrix} \theta_1(-{\bm k})  \\
                                                                                                                                          \pi_2(-{\bm k}) \\
                                                                                                                                        \end{pmatrix}
\nonumber\\
&& + \frac{1}{2V}\sum_{\bm k} \left(\theta_2({\bm k}),\pi_1({\bm k})\right)\begin{pmatrix} -w_2 m_0^2 + a k^2  &  -w_2 n_0 m_0  \\
                                                                                                                                         -w_2 n_0 m_0           &  h/m_0 - w_2 n_0^2 + c k^2 \\
                                                                                                                                         \end{pmatrix}
                                                                                                                 \begin{pmatrix} \theta_2(-{\bm k})  \\
                                                                                                                                          \pi_1(-{\bm k}) \\
                                                                                                                                        \end{pmatrix}
\nonumber\\
&& + \frac{1}{2V}\sum_{\bm k} \left(\theta_3({\bm k}),\pi_3({\bm k})\right)\begin{pmatrix} u n_0^2 + a k^2  &  (w_1+w_2) n_0 m_0  \\
                                                                                                                                         (w_1+w_2) n_0 m_0   &  h/m_0 - w_2 n_0^2 + c k^2 \\
                                                                                                                                         \end{pmatrix}
                                                                                                                 \begin{pmatrix} \theta_3(-{\bm k})  \\
                                                                                                                                          \pi_3(-{\bm k}) \\
                                                                                                                                        \end{pmatrix}\ .
\label{eq:2.16}
\eea 
\end{widetext}                                                                                                                                       
All eigenvalues are positive
provided Eq.~(\ref{eq:2.9}) holds, and all of them remain positive for $k=0$. Hence there are no Goldstone modes. 
This reflects the fact that the field polarizes the homogeneous magnetization via the Zeeman term, which in turn 
polarizes the staggered magnetization via the $w_2$ term, so any deviation from the collinear field configuration costs energy. 

 \medskip
 
In the limit of a vanishing field, $h = m_0 = 0$, two of the positive eigenvalues in Case 2 vanish at $k=0$, and Case 1 yields one
additional zero eigenvalue. This reflects the two zero-field AFM Goldstone modes that are represented
by the transverse fluctuations of the staggered magnetization:
\bse
\label{eqs:2.17}
\be
g_{1,2}({\bm k}) = \theta_{1,2}({\bm k})
\label{eq:2.17a}
\ee
with susceptibilities
\be
\chi_{g_1}(k) = \chi_{g_2}(k) = 1/a\,k^2\ .
\label{eq:2.17b}
\ee
\ese

\medskip

\paragraph{b. Coexisting AFM and homogeneous order}
\label{par:II.A.3.b}
$ $
\medskip\par\noindent
{\it Case 1:} $w_2 > 0$ (orthogonal-order-parameters case)
\par
This case is obtained from the transverse-field case, Eq.~(\ref{eq:2.14}), by taking the limit $h\to 0$ at fixed $m_0$. The eigenvalue that corresponds
to the $\pi_2$ fluctuations now vanishes in the $k\to 0$ limit in addition to the one that corresponds to the $\theta_1$
fluctuations, and the $(\theta_2,\pi_3)$ system contributes a third zero eigenvalue. We thus have three Goldstone modes, namely
\bse
\label{eqs:2.18}
\bea
g_1({\bm k}) &=& \theta_1({\bm k})\ ,
\nonumber\\
g_2({\bm k}) &=& \pi_2({\bm k})\ ,
\nonumber\\
g_3({\bm k}) &=& \theta_2({\bm k}) + i\pi_3({\bm k})\ ,
\label{eq:2.18a}
\eea
with susceptibilities
\bea
\chi_{g_1}(k) &=& 1/a\,k^2\ ,
\nonumber\\
\chi_{g_2}(k) &=& 1/c\,k^2\ ,
\nonumber\\
\chi_{g_3}(k) &=& \frac{n_0^2 + m_0^2}{n_0^2 c + m_0^2 a}\,\frac{1}{k^2}\ .
\label{eq:2.18}
\eea
\ese
Physically, the situation is as follows. In the absence of a coupling between ${\bm n}$ and ${\bm m}$ the transverse fluctuations of both
order parameters would be soft. However, the $w_2$ coupling enforces the condition that the two order parameters are orthogonal. 
This leads to one constraint, which reduces the number of soft modes from four to three. 

\medskip\par\noindent
{\it Case 2:} $w_2 < 0$ (collinear-order-parameters case)
\par
This case is obtained from the longitudinal-field case, Eq.~(\ref{eq:2.16}), by taking the limit $h\to 0$ at fixed $m_0$. 
Of the six eigenvalues, two vanish in this limit at $k=0$. There are two Goldstone modes,
\bea
g_1({\bm k}) &=& \theta_1({\bm k}) + i\pi_2({\bm k})\ ,
\nonumber\\
g_2({\bm k}) &=& \theta_2({\bm k}) + i\pi_1({\bm k})\ ,
\label{eq:2.19}
\eea
with susceptibilities
\be
\chi_{g_1}(k) = \chi_{g_2}(k) = \frac{n_0^2 + m_0^2}{n_0^2 c + m_0^2 a}\,\frac{1}{k^2}\ .
\label{eq:2.20}
\ee
Physically, the homogeneous magnetization gets slaved to the staggered magnetization by the $w_2$ coupling, and the soft-mode
structure is that of an antiferromagnet.

\subsection{Dynamics}
\label{subsec:II.B}

The dynamics of the order-parameter fields ${\bm n}$ and ${\bm m}$ are governed by the basic equation of motion
that describes the precession of the magnetic moment ${\bm M}$ in an effective magnetic field,\cite{Landau_Lifshitz_1935, Ma_1976}
\be
\partial_t {\bm M}({\bm x},t) = {\bm M}({\bm x},t)\times\frac{\delta F}{\delta{\bm M}({\bm x})}\Bigr\vert_{{\bm M}({\bm x},t)}\ .
\label{eq:2.21}
\ee
We put the gyromagnetic ratio equal to unity, which amounts to measuring the magnetization in units of the magnetic moment.
By using
\bea
\frac{\delta F}{\delta M_i({\bm x})} &=& \int d{\bm y}\,\left(\frac{\delta F}{\delta m_j({\bm y})}\,\frac{\delta m_j({\bm y})}{\delta M_i({\bm x})} 
     + \frac{\delta F}{\delta \nu_j({\bm y})}\,\frac{\delta \nu_j({\bm y})}{\delta M_i({\bm x})} \right)
\nonumber\\
 &=& \frac{\delta F}{\delta m_i({\bm x})} + \frac{\delta F}{\delta \nu_i({\bm x})} 
\label{eq:2.22}
\eea
and coarse-graining Eq.~(\ref{eq:2.21}) we obtain
\bse
\label{eqs:2.23}
\be
\partial_t {\bm m}(x) = {\bm m}(x)\times\frac{\delta F}{\delta{\bm m}({\bm x})}\Biggr\vert_{{\bm m}(x)}
   \hskip -10pt + {\bm n}({\bm x},t)\times\frac{\delta F}{\delta{\bm n}({\bm x})}\Biggr\vert_{{\bm n}(x)} ,
\label{eq:2.23a}
\ee
where $x = ({\bm x},t)$. Multiplying Eq.~(\ref{eq:2.21}) by $f({\bm x})$ and coarse-graining we obtain
\be
\partial_t{\bm n}(x) = {\bm n}(x)\times\frac{\delta F}{\delta{\bm m}({\bm x})}\Biggr\vert_{{\bm m}(x)} 
   \hskip -10pt +{\bm m}(x)\times\frac{\delta F}{\delta{\bm n}({\bm x})}\Biggr\vert_{{\bm n}(x)} ,
\label{eq:2.23b}
\ee
\ese
where we have used Eq.~(\ref{eq:2.3.c}).  Adding dissipative terms we finally 
obtain \cite{Freedman_Mazenko_1976, Freedman_Mazenko_footnote}
\bse
\label{eqs:2.24}
\bea
\partial_t\,{\bm n} &=& -\Gamma_0\,\frac{\delta F}{\delta{\bm n}} + {\bm n}\times\frac{\delta F}{\delta{\bm m}} 
                                  + {\bm m}\times\frac{\delta F}{\delta{\bm n}}\ ,
\label{eq:2.24a}\\
\partial_t\,{\bm m} &=& \lambda_0\,{\bm\nabla}^2\,\frac{\delta F}{\delta{\bm m}} + {\bm n}\times\frac{\delta F}{\delta{\bm n}} 
      + {\bm m}\times\frac{\delta F}{\delta{\bm m}}\ ,\qquad
\label{eq:2.24b}
\eea
\ese
where $\Gamma_0$ and $\lambda_0$ are bare kinetic coefficients. The functional form of the dissipative term in Eq.~(\ref{eq:2.24a})
(constant $\Gamma_0$) reflects the fact that the staggered magnetization ${\bm n}$ is not a conserved quantity. The gradient-squared form of the corresponding
term in Eq.~(\ref{eq:2.24b}) is valid provided the total magnetization is conserved. If it is not, e.g., due to the presence of magnetic impurities,
then this term must also have a constant coefficient and we have, instead of Eq.~(\ref{eq:2.24b}),
\be
\partial_t\,{\bm m} = -\mu_0\,\frac{\delta F}{\delta{\bm m}} + {\bm n}\times\frac{\delta F}{\delta{\bm n}} 
      + {\bm m}\times\frac{\delta F}{\delta{\bm m}}\ .
      \tag{2.24b'}
\ee
The last term on the right-hand-side of Eq.~(\ref{eq:2.24a}) is often omitted since it is irrelevant for the critical dynamics
of a classical antiferromagnet.\cite{Freedman_Mazenko_1976} Equations~(\ref{eqs:2.24}) without this term, and with $v = w_1 = w_2 = 0$
in Eq.~(\ref{eq:2.4}), is often referred to as Model G in the classification of Ref.~\onlinecite{Hohenberg_Halperin_1977}. We also note
that in order to calculate correlation functions one needs to add Langevin forces on the right-hand-sides of Eqs.~(\ref{eqs:2.24}), see
Appendix~\ref{app:B}.
Alternatively, one can calculate response functions in the presence of the fields ${\bm h}$ and $\textswab{h}$. We will take the latter
approach; the fluctuation-dissipation theorem can then be used to determine the correlation functions.

\section{Linearized equations of motion, and spin excitations}
\label{sec:III}

We now parameterize the fields ${\bm n}$ and ${\bm m}$ as in Eq.~(\ref{eq:2.13})
and linearize the kinetic equations (\ref{eqs:2.24}) in the small fluctuations $\theta_{1,2,3}$ and $\pi_{1,2,3}$.
This yields a $6\times 6$ system of linear equations. The solutions for ${\textswab h}=0$ give the eigenoscillations of the
antiferromagnet. For counting purposes we will treat this as analogous to a mechanical system; that is, we have six
eigenvalues and corresponding eigenvectors that characterize six modes. In the case of propagating modes, pairs of
modes that propagate in opposite directions form one spin wave.

\bigskip 
\subsection{Conserved homogeneous magnetization; AFM order only}
\label{subsec:III.A}

\subsubsection{Zero field}
\label{subsubsec:III.A.1}

For completeness, we first recall the well-known results for a vanishing external field, ${\bm h}=0$. The six equations decouple
into two identical pairs of $2\times 2$ systems for $\theta_1,\pi_1$ and $\theta_2,\pi_2$, respectively, and two single equations for
$\theta_3$ and $\pi_3$, respectively. They read
\be
\begin{pmatrix} i\Omega + \Gamma_0 a k^2  &  -n_0(t + w_1 n_0^2 + c k^2)  \\
                         n_0 a k^2                               &  i\Omega + \lambda_0(t + w_1 n_0^2 + c k^2) k^2 \\
                         \end{pmatrix}
\begin{pmatrix} \theta_{1,2} \\
                         \pi_{1,2}     \\
                         \end{pmatrix} = 0 \ .
\label{eq:3.1}
\ee  
and
\bea
\left[ i\Omega + \Gamma_0(2 u n_0^2 + a k^2)\right] \theta_3 &=& 0\ ,\qquad
\label{eq:3.2}\\
\left[i\Omega + \lambda_0\left(t + (w_1 + w_2)n_0^2 + c k^2)\right)k^2\right] \pi_3 &=& 0\ .
\label{eq:3.3}
\eea
Equation (\ref{eq:3.1}) yields two identical pairs of gapless propagating modes with eigenfrequencies
\bse
\label{eqs:3.4}
\be
\Omega_{1,\pm} = \Omega_{2,\pm} = \pm c_1\,k + \frac{i}{2}\,\Gamma_1 a k^2 + O(k^3)\ .
\label{eq:3.4a}
\ee
The speed of the propagating modes is
\be
c_1 = n_0\sqrt{a} \sqrt{t + w_1 n_0^2} = n_0 \sqrt{a h/m_0}\bigr|_{h\to 0}
\label{eq:3.4b}
\ee
and the damping coefficient is
\be
\Gamma_1 = \Gamma_0 + \lambda_0 (c_1/n_0 a)^2\ .
\label{eq:3.4c}
\ee
The right eigenvectors are
\be
{\bm\nu}_{{\rm R}1,\pm} \equiv \begin{pmatrix} \theta_1 \\
                                                                           \pi_1      \\
                                                                           \end{pmatrix}_{\pm}
                       = \begin{pmatrix} 1 \\
                                                    \pm i(n_0 a/c_1) k + O(k^2) \\
                                                   \end{pmatrix}
\label{eq:3.4d}
\ee
and ${\bm\nu}_{{\rm R}2,\pm} = {\bm\nu}_{{\rm R}1,\pm}$, so the long-wavelength spin waves are
transverse $\theta$-fluctuations with a small admixture of $\pi$-fluctuations.
Also of interest are the left eigenvectors
\be
{\bm\nu}_{{\rm L}1,\pm} \equiv (\theta_1,\pi_1)_{\pm} = \left(\pm i (n_0 a/c_1)k + O(k^2),1\right)
\label{eq:3.4e}
\ee
\ese
and ${\bm\nu}_{{\rm L}2,\pm} = {\bm\nu}_{{\rm L}1,\pm}$. Note that the left eigenvectors are structurally
very different from the right ones. This is important for calculating time correlation functions, and for
ensuring that the fluctuation-dissipation theorem holds, see Appendix~\ref{app:B}.

In addition, there is a pure $\pi_3$ mode described by Eq.~(\ref{eq:3.3}) which is diffusive with an eigenfrequency
\bse
\label{eqs:3.5}
\be
\Omega_3 = i D k^2 + O(k^4)\ ,
\label{eq:3.5a}
\ee
and a diffusion constant
\be
D = \lambda_0\left(t + (w_1+w_2)n_0^2\right)\ .
\label{eq:3.5b}
\ee
\ese
Finally, Eq.~(\ref{eq:3.2}) describes a pure $\theta_3$ mode which is relaxational with eigenfrequency
\be
\Omega_4 = i\,2 \Gamma_0 u\,n_0^2 + O(k^2)\ .
\label{eq:3.6}
\ee

These are the well-known results for an isotropic Heisenberg antiferromagnet: There are two spin waves with
a linear dispersion relation and quadratic damping, and the dynamics of the longitudinal homogeneous 
magnetization are diffusive.

Equation (\ref{eq:3.6}) implies that the longitudinal response function in the limit of long wavelengths and
low frequencies is a constant. This result requires a qualification, as it changes qualitatively if one goes 
beyond the linearized theory: The coupling between the longitudinal and transverse degrees of freedom
leads to a longitudinal response function that diverges, in classical antiferromagnets, as $1/k^{d-4}$ in
dimensions $d<4$,\cite{Vaks_Larkin_Pikin_1967, Brezin_Wallace_1973} and as $1/k^{d-3}$ at $T=0$
in dimensions $d<3$.\cite{Bharadwaj_Belitz_Kirkpatrick_2016} This is most easily seen by employing
a nonlinear sigma model; alternatively, one can apply the renormalization-group techniques reviewed
in Ref.~\onlinecite{Hohenberg_Halperin_1977} to the present formalism.

\subsubsection{Transverse field}
\label{subsubsec:III.A.2}

In Sec.~\ref{subsec:II.A} we saw that if $w_2>0$ and ${\bm h}\neq 0$, the lowest free-energy state is
a configuration where ${\bm m}_0$ and ${\bm h}$ are collinear and perpendicular to ${\bm n}_0$.
We choose ${\bm n}_0 = (0,0,n_0)$, ${\bm m}_0 = (m_0,0,0)$, and ${\bm h} = (h,0,0)$. The linearized
kinetic equations then take the form of two $3\times 3$ systems for $(\theta_1,\theta_3,\pi_1)$ and
$(\theta_2,\pi_2,\pi_3)$, respectively. They read
\begin{widetext}
\be
\begin{pmatrix} i\Omega + \Gamma_0 a k^2 & 2n_0^2 m_0(u-w_1)+m_0ak^2                  & -n_0\left(h/m_0 + 2m_0^2(v-w_1) + c\,k^2\right)                    \\
                         -m_0 a k^2                            &  i\Omega + \Gamma_0(2un_0^2 + a k^2) &  2\Gamma_0 w_1 n_0 m_0                                                     \\
                          n_0 a k^2                             &  2\lambda_0 w_1 n_0 m_0 k^2                 &  i\Omega + \lambda_0\left(h/m_0 + 2vm_0^2 + ck^2\right)k^2 \\            
                                                                       \end{pmatrix}
\begin{pmatrix} \theta_1 \\
                          \theta_3 \\
                          \pi_1
                          \end{pmatrix} = 0\ ,
\label{eq:3.7}                          
\ee
and
\be
\begin{pmatrix} i\Omega + \Gamma_0 (w_2m_0^2 + a k^2) & -n_0 (h/m_0 + c\,k^2)                                  & -\Gamma_0 w_2 n_0 m_0                            \\
                         n_0 a k^2                                                      &  i\Omega + \lambda_0 (h/m_0 + c\,k^2)k^2 &  h + m_0 c\,k^2                                             \\
                         -\lambda_0 w_2 n_0 m_0 k^2                       &  -(h + m_0 c\,k^2)                   &  i\Omega + \lambda_0(h/m_0 + w_2 n_0^2 + c\,k^2)k^2   \\            
                                                                       \end{pmatrix}
\begin{pmatrix} \theta_2 \\
                          \pi_2     \\
                          \pi_3
                          \end{pmatrix} = 0\ .
\label{eq:3.8}                          
\ee\end{widetext}

The former yields one pair of gapless propagating modes with eigenfrequencies
\bse
\label{eqs:3.9}
\be
\Omega_{1,\pm} = \pm c_1\,k + \frac{i}{2}\,\Gamma_1 a k^2 + O(k^3)\ ,
\label{eq:3.9a}
\ee
and right and left eigenvectors
\be
{\bm\nu}_{{\rm R}1,\pm} \equiv \begin{pmatrix} \theta_1 \\
                                                                           \theta_3 \\
                                                                           \pi_1 \\ \end{pmatrix}_{{\rm R}1,{\pm}} = \begin{pmatrix} 1 \\ 
                                                                                                                                                                    \mp i\,\frac{a w_1 m_0}{u\,c_1}\,k + O(k^2) \\
                                                                                                                                                                    \pm i\,\frac{n_0\,a}{c_1}\,k + O(k^2) \\
                                                                                                                                                                    \end{pmatrix}\ ,
\label{eq:3.9b}
\ee
\bea
{\bm\nu}_{{\rm L}1,\pm} &\equiv& (\theta_1,\theta_3,\pi_1)_{\rm{L}1,\pm} 
\nonumber\\
 &=& \left(\pm i\,\frac{n_0 a}{c_1}\,k, \mp i\,\frac{n_0 m_0 a}{\Gamma_0 c_1}
     \left(1 - \frac{w_1}{u}\right) k ,1\right)
     \nonumber\\
     &&\hskip 80pt + O(k^2)
\label{eq:3.9c}
\eea
This is a generalization of one of the two transverse modes from Eqs.~(\ref{eqs:3.4}). The magnetic field
leads to a $\theta_3$-component of the eigenoscillation, and it modifies the speed and the damping
coefficient of the mode:
\bea
c_1 &=& n_0 \sqrt{a} \sqrt{h/m_0 + 2(v -w_1^2/u)m_0^2}\ ,
\label{eq:3.9d}\\
\Gamma_1 &=& \Gamma_0 + \lambda_0\,(c_1/n_0\,a)^2 + \frac{m_0^2}{\Gamma_0}\,(1 - w_1/u)^2\ .
\nonumber\\
\label{eq:3.9e}
\eea
\ese
For $h=0$ these expressions correctly reduce to Eqs.~(\ref{eqs:3.4}). They hold in the limit of asymptotically
small $k$ for fixed damping coefficients $\Gamma_0$, $\lambda_0$, i.e., for $k \ll c_1/a\Gamma_1$. 
The limits $k\to 0$ and $\Gamma_0, \lambda_0 \to 0$ do not commute, as is obvious from the last term in Eq.~(\ref{eq:3.9e}). 
In the limit of vanishing damping coefficients at fixed small $k$ the modes remain propagating and gapless, but
the speed of the propagation changes. One finds for the eigenfrequencies
\bse
\label{eqs:3.10}
\be
\tilde\Omega_{1,\pm} = \pm \tilde c_1\,k + O(k^2, \Gamma_0, \lambda_0)\ ,
\label{eq:3.10a}
\ee
and for the right and left eigenvectors
\be
{\tilde{\bm\nu}}_{{\rm R}1,{\pm}} = \begin{pmatrix} 1 \\ 
                                                                                \mp i\,(a\, m_0/\tilde c_1) k + O(k^2) \\
                                                                                \pm i\,(a\, n_0/\tilde c_1) k + O(k^2) \\
                                                                                \end{pmatrix}
\label{eq:3.10b}
\ee
\bea
{\tilde{\bm\nu}}_{{\rm L}1,{\pm}} &=& \Bigl(\pm i c_1 k, -2 n_0^2 m_0(u-w_1), 
\nonumber\\
&& \quad n_0(\frac{h}{m_0} + 2m_0^2(v-w_1))\Bigr) + O(k^2,\Gamma_0,\lambda_0)
\nonumber\\
\label{eq:3.10c}
\eea
with
\be
\tilde c_1 = n_0 \sqrt{a} \sqrt{h/m_0 + 2(u + v - 2w_1)m_0^2}\ ,
\label{eq:3.10d}
\ee
\ese

The second transverse modes are gapped propagating modes with eigenfrequencies
\bse
\label{eqs:3.11}
\be
\Omega_{2,\pm} = \pm\sqrt{h^2 + c_2^2 k^2}  + \frac{i}{2}\,\Gamma_2(h,k) a k^2 + O(\epsilon^3)\quad
\label{eq:3.11a}
\ee
and right and left eigenvectors
\be
{\bm\nu}_{{\rm R}2,\pm} \equiv \begin{pmatrix} \theta_2  \\
                                                                           \pi_2  \\
                                                                           \pi_3 \\
                                                                           \end{pmatrix}_{{\rm R}2,\pm} 
          = \begin{pmatrix} 1  \\
                                       \pm i\,\frac{m_0/h}{n_0}\,\sqrt{h^2+ c_2^2 k^2} + O(\epsilon^2)\\
                                        m_0/n_0 + O(\epsilon^2)\\
                                        \end{pmatrix},
\label{eq:3.11b}
\ee
\bea
{\bm\nu}_{{\rm L}2,\pm} &\equiv& \left(\theta_2,\pi_2,\pi_3\right)_{\rm{L}2,\pm} 
\nonumber\\
&=& \left(\frac{n_0 a k^2}{h} + O(\epsilon^2), \frac{\mp i}{h}\,\sqrt{h^2 + c_2^2 k^2} + O(\epsilon^2), 1\right).
\nonumber\\
\label{eq:3.11c}
\eea
where
 $\epsilon = O(k,h)$. Here
\be
c_2 = n_0\sqrt{a\,h/m_0}\ ,
\label{eq:3.11d}
\ee
and 
\be
\Gamma_2(h,k) = \frac{c_2^2/a}{h^2 + c_2^2\,k^2}\,\left[\Gamma_0 a k^2 + \lambda_0 (2 h^2 + c_2^2 k^2)/n_0^2\,a\right]\ .
\label{eq:3.11e}
\ee
\ese
We see that in this mode the magnetic field opens a gap of magnitude $h$. For $h=0$ we again recover the expressions in
Eqs.~(\ref{eqs:3.4}). Note that $\pi_3$ is part of the linear combination that comprises the gapped propagating modes.
For $h=0$ it decouples and is diffusive, see Eqs.~(\ref{eqs:3.5}). 

In addition to these propagating modes there are two relaxational modes. One has an eigenfrequency
\bse
\label{eqs:3.12}
\be
\Omega_3 = i 2\Gamma_0 u n_0^2 + O(k^2)
\label{eq:3.12a}
\ee
and right and left eigenvectors
\be
{\bm\nu}_{\rm{R}3} \equiv \begin{pmatrix} \theta_1   \\
                                                      \theta_3  \\
                                                       \pi_1  \\
                                                       \end{pmatrix}_{\rm{R}3} = \begin{pmatrix} (1-w_1/u) m_0/\Gamma_0 + O(k^2) \\
                                                                                                                           1 \\  
                                                                                                                           O(k^2) \\ \end{pmatrix}\ ,
\label{eq:3.12b}
\ee
\be
{\bm\nu}_{\rm{L}3} \equiv  \left(\theta_1, \theta_3, \pi_1 \right)_{\rm{L}3} = \left(O(k^2),1,\frac{w_1 m_0}{u n_0}\right) \ .
\label{eq:3.12c}
\ee
\ese
This is valid in the limit of asymptotically small wavenumber. In the limit of asymptotically small damping one finds instead
\bse
\label{eqs:3.13}
\be
\tilde\Omega_3 = i 2\Gamma_0 u n_0^2 c_1^2/\tilde c_1^2
\label{eq:3.13.a}
\ee
and
\be
\tilde{\bm\nu}_{\rm{R}3} = \begin{pmatrix}  0  \\
                                                                    1  \\
                                                                    \frac{2n_0^2 m_0(u-w_1) + m_0 a k^2}{n_0 h/m_0 + 2 n_0 m_0^2(v-w_1) + n_0 c k^2}
                                                                    \end{pmatrix}
+ O(\Gamma_0,\lambda_0)\ .
\label{eq:3.13.b}
\ee
\ese
The other relaxational mode has an eigenfrequency
\bse
\label{eqs:3.14}
\be
\Omega_4 = i \Gamma_0 w_2 m_0^2 + O(k^2)
\label{eq:3.14a}
\ee
and right and left eigenvectors
\be
{\bm\nu}_{\rm{R}4} \equiv \begin{pmatrix} \theta_2  \\
                                                                   \pi_2   \\
                                                                   \pi_3 \\
                                                                   \end{pmatrix}_{\rm{R}4} = \begin{pmatrix} 1  \\
                                                                                                                                       O(k^2) \\
                                                                                                                                       O(k^2) \\
                                                                                                                                       \end{pmatrix}\ ,
\label{eq:3.14b}
\ee
\be
{\bm\nu}_{\rm{L}4} \equiv \left(\theta_2,\pi_2,\pi_3\right)_{\rm{L}4} = \left(1,O(k^2),-n_0/m_0 + O(k^2)\right) \ .
\label{eq:3.14c}
\ee
\ese

\subsubsection{Longitudinal field}
\label{subsubsec:III.A.3}

If $w_2 < 0$, then in a nonzero external field the lowest free-energy configuration is one where ${\bm m}_0$,
${\bm h}$, and ${\bm n_0}$ are all collinear; we choose them to be parallel to $(0,0,1)$. The linearized kinetic
equations then decouple into one $2\times 2$ system for $(\theta_3,\pi_3)$, 
\begin{widetext}
\be
\begin{pmatrix} i\Omega + \Gamma_0 (2 u n_0^2 + a k^2) & 2(w_1+w_2)\Gamma_0 n_0 m_0                       \\
                         2(w_1+w_2)\lambda_0 n_0 m_0 k^2       & i\Omega + \lambda_0(h/m_0 + 2 v m_0^2 + c k^2)k^2   \\         
                                                                       \end{pmatrix}
\begin{pmatrix} \theta_3 \\
                          \pi_3 \\
                          \end{pmatrix} = 0\ ,
\label{eq:3.15}                          
\ee
and one $4\times 4$ system for $(\theta_1,\theta_2,\pi_1,\pi_2)$, 
\bse
\label{eqs:3.16}
\bea
\begin{pmatrix} i\Omega + \Gamma_2(k)  &  \tilde m_0(k) &  -\tilde n_0(k)  &  w_2 \Gamma_0 n_0 m_0   \\
                         -\tilde m_0(k)  & i\Omega + \Gamma_2(k)  &  -w_2\Gamma_0 n_0 m_0  &  -\tilde n_0(k) \\     
                         n_0 ak^2  &  -w_2 \lambda_0 n_0 m_0 k^2  &  i\Omega + \tilde\lambda_0(k) k^2  &  h + m_0 ck^2                                 \\
                         w_2 \lambda_0 n_0 m_0 k^2  &  n_0 ak^2  &  -(h + m_0 ck^2) & i\Omega + \tilde\lambda_0(k) k^2                                          \\    
                                                                       \end{pmatrix}
                                                                       \begin{pmatrix} \theta_1 \\
                          \theta_2 \\
                          \pi_1 \\
                          \pi_2 \\
                          \end{pmatrix} = 0\ .\hskip 60pt
\label{eq:3.16a}                          
\eea
Here we have defined
\bea
\Gamma_2(k) &=& \Gamma_0 (-w_2 m_0^2 + a k^2)\ ,
\label{eq:3.16b}\\
\tilde m_0(k) &=& m_0\left( w_2(n_0^2 - m_0^2) + ak^2\right) 
\label{eq:3.16c}\\
\tilde n_0(k) &=& n_0\left(h/m_0 - w_2 (n_0^2-m_0^2) + c k^2\right) 
\label{eq:3.16d}\\
\tilde\lambda_0(k) &=& \lambda_0(h/m_0 - w_2 n_0^2 + ck^2)
\label{eq:3.16e}
\eea
\ese
Equations (\ref{eqs:3.16}) lead to two pairs of gapped propagating modes. For small $h$ and $k$ we find for the eigenfrequencies\cite{4_by_4_footnote}
\bse
\label{eqs:3.17}
\bea
\Omega_{1,\pm} &=& \frac{\pm 1}{\sqrt{2}}\left[(1 + w_2^2 n_0^4 m_0^2/h^2) h^2 + 2 {\tilde c}^2 k^2
   - \sqrt{(1 - w_2^2 n_0^4 m_0^2/h^2)^2 h^4 + 4 h^2 (1 + w_2 n_0^2 m_0/h)^2 {\tilde c}^2 k^2} + O(\epsilon^4)\right]^{1/2} 
\nonumber\\
&& \hskip 270pt -iw_2\Gamma_0m_0^2 \left[1 + O(k^2/m_0^2)\right]\ ,
\label{eq:3.17a}\\
\Omega_{2,\pm} &=& \frac{\pm 1}{\sqrt{2}} \left[(1 + w_2^2 n_0^4 m_0^2/h^2) h^2 + 2 {\tilde c}^2 k^2
   + \sqrt{(1 - w_2^2 n_0^4 m_0^2/h^2)^2 h^4 + 4 h^2 (1 + w_2 n_0^2 m_0/h)^2 {\tilde c}^2 k^2} + O(\epsilon^4)\right]^{1/2} 
\nonumber\\
&& \hskip 250pt +\, i \lambda_0 (h/m_0)\,k^2  \left[1 + O(k^2/m_0^2)\right]\ ,
\label{eq:3.17b}
\eea
where the innermost square root is defined as $\sqrt{x^2} = x$ irrespective of the sign of $x$, 
\be
{\tilde c} = n_0 \sqrt{a}\sqrt{h/m_0 - w_2 n_0^2}\
\label{eq:3.17c}
\ee
and $\epsilon$ can stand for either $h$ or $k$. The damping coefficients are easily obtained to $O(\epsilon^2)$, but
the results are complicated and we show only the leading terms for $k\to 0$ at fixed $h$.
The corresponding right and left eigenvectors are, at $k=0$,
\bea
{\bm\nu}_{{\rm R}1,\pm} \equiv \begin{pmatrix} \theta_1 \\
                                                                           \theta_2 \\
                                                                           \pi_1  \\
                                                                           \pi_2 \\
                                                                           \end{pmatrix}_{\rm{R}1\pm} = \begin{pmatrix} 1  \\
                                                                                                                                                    \mp i \\
                                                                                                                                                    0  \\
                                                                                                                                                    0  \\
                                                                                                                                                    \end{pmatrix}\ ,
\label{eq:3.17d}\\
{\bm\nu}_{{\rm R}2,\pm} \equiv \begin{pmatrix} \theta_1 \\
                                                                           \theta_2 \\
                                                                           \pi_1  \\
                                                                           \pi_2 \\
                                                                           \end{pmatrix}_{\rm{R}2\pm} = \begin{pmatrix}  1 \\
                                                                                                                                                     \mp i  \\
                                                                                                                                                     \pm i m_0/n_0  \\
                                                                                                                                                     m_0/n_0 \\
                                                                                                                                                     \end{pmatrix}\ ,
\label{eq:3.17e}\\
{\bm\nu}_{{\rm L}1,\pm} \equiv  \left(\theta_1, \theta_2, \pi_1, \pi_2\right)_{{\rm L}1,\pm} = \left(\mp i m_0/n_0,m_0/n_0,1,\pm i\right)\ ,
\label{eq:3.17f}\\
{\bm\nu}_{{\rm L}2,\pm} \equiv  \left(\theta_1, \theta_2, \pi_1, \pi_2\right)_{{\rm L}2,\pm} = (0,0,1,\pm i)\ .
\label{eq:3.17g}
\eea
\ese
\end{widetext}
Note that the first two modes come with a damping coefficient that does not vanish as $k\to 0$, whereas the other two have a
damping coefficient that vanishes as $k^2$, as in the transverse-field case. Also note
that if the coupling constant $w_2$ were neglected, the first pair of modes would be gapless with a quadratic
dispersion relation. Keeping all coupling constants consistent with the symmetry of the problem is thus important for obtaining the
correct soft-mode structure. 

From the $2\times 2$ system we obtain a diffusive mode with eigenfrequency
\bse
\label{eqs:3.18}
\be
\Omega_3 = i D_3 k^2\ .
\label{eq:3.18a}
\ee
The diffusion constant is given by
\be
D_3 = \lambda_0 \left[\frac{h}{m_0} + 2 v m_0^2 - 2(w_1+w_2)^2m_0^2/u\right]\ ,
\label{eq:3.18b}
\ee
and the right and left eigenvectors are
\be
{\bm\nu}_{\rm{R}3} \equiv \begin{pmatrix} \theta_3  \\
                                                                   \pi_3  \\
                                                                   \end{pmatrix}_{\rm{R}3} = \begin{pmatrix} \frac{-(w_1 + w_2)}{u}\,\frac{m_0}{n_0}  + O(k^2) \\
                                                                                                                                      1 \\
                                                                                                                                      \end{pmatrix} \ ,
\label{eq:3.18c}
\ee
\be
{\bm\nu}_{\rm{L}3} \equiv \left(\theta_3,\pi_3\right)_{\rm{L}3} = \left( \frac{-(w_1+w_2)\lambda_0 m_0}{u\Gamma_0 n_0}\,k^2 + O(k^4),1\right) \ .
\label{eq:3.18d}
\ee
\ese
This is a generalization of the diffusive $\pi_3$ mode in zero field. 

Finally, there is a relaxational mode with eigenfrequency
\bse
\label{eqs:3.19}
\be
\Omega_4 = i 2\Gamma_0 u n_0^2  + O(k^2)
\label{eq:3.19a}
\ee
and right and left eigenvectors
\bea
{\bm\nu}_{\rm{R}4} &\equiv& \begin{pmatrix} \theta_3  \\
                                                                   \pi_3  \\
                                                                   \end{pmatrix}_{\rm{R}4} = \begin{pmatrix} 1 \\
                                                                                                                                      \frac{(w_1+w_2)\lambda_0 m_0}{u\Gamma_0 n_0}\,k^2 + O(k^4) \\
                                                                                                                                      \end{pmatrix} \ ,
\label{eq:3.19b}\\
{\bm\nu}_{\rm{L}4} &\equiv& \left(\theta_3,\pi_3\right)_{\rm{L}4} = \left(1,\frac{w_1+w_2}{u}\,\frac{m_0}{n_0} + O(k^2)\right).\qquad\quad
\label{eq:3.19c}
\eea
\ese
which generalizes the relaxational $\theta_3$ mode in zero field.

\subsection{Conserved homogeneous magnetization; coexisting AFM and homogeneous orders}
\label{subsec:III.B}

We now consider the modes for the case of coexisting AFM and homogeneous orders. As discussed in Sec.~\ref{subsubsec:II.A.2}
we need to distinguish again between $w_2>0$, which leads to orthogonal order parameters, and $w_2<0$, which leads to
collinear order parameters.

\subsubsection{Orthogonal order parameters}
\label{subsubsec:III.B.1}

The gapless propagating modes are obtained by taking the straightforward limit $h\to 0$ at fixed $m_0$ in Eqs.~(\ref{eqs:3.9}).
The results from Sec.~\ref{subsubsec:III.A.2} remain valid except that the speed of the modes now is
\bse
\label{eqs:3.20}
\be
c_1 = n_0 \sqrt{2a(v - w_1^2/u)m_0^2}
\label{eq:3.20a}
\ee
in the limit of vanishing wave number, and
\be
\tilde c_1 = n_0 \sqrt{2a(u+v-2w_1)m_0^2}
\label{eq:3.20b}
\ee
in the limit of vanishing damping coefficients.
\ese

For the other oscillating modes the result cannot simply be read off from the results of Sec.~\ref{subsubsec:III.A.2},
for two reasons: The oscillation frequency vanishes to $O(k)$ in the limit $h\to 0$ at fixed $m_0$, and in the damping
term the limits $h\to 0$ and $\Gamma_0\to 0$ do not commute in the coexisting-orders case. An analysis of
Eq.~(\ref{eq:3.8}) yields
\bse
\label{eqs:3.21}
\be
\Omega_{2,\pm} = \pm d_2 k^2\left(1 - \frac{a^2 c n_0^2}{2 m_0^2 w_2 d_2^2}\,k^2\right) + \frac{i}{2}\lambda_2 c k^4 + O(k^6)
\label{eq:3.21a}
\ee
in the limit of asymptotically small wave numbers. Here
\be
d_2 = \sqrt{m_0^2 c^2 + n_0^2 a c}
\label{eq:3.21b}
\ee
and
\be
\lambda_2 = \lambda_0 (2 + n_0^2 a/m_0^2 c)\ .
\label{eq:3.21c}
\ee
The corresponding right and left eigenvectors are
\bea
{\bm\nu}_{R2,\pm} &\equiv& \begin{pmatrix} \theta_2 \\
                                                             \pi_2      \\
                                                             \pi_3 \\
                                                             \end{pmatrix}_{R2,\pm} \hskip -10pt = 
                                                             \begin{pmatrix}
                                                             1 \\
                                                             \pm i d_2/n_0 c +O(k^2)\\
                                                             m_0/n_0 + a k^2/w_2 n_0 m_0 + O(k^4)\\
                                                             \end{pmatrix}.
\nonumber\\                                                             
\label{eq:3.21d}\\
{\bm\nu}_{L2,\pm} &\equiv& \left(\frac{\pm i a c n_0^2/d_2 + \lambda_0 w_2 n_0^2}{\Gamma_0 w_2 n_0 m_0}\,k^2, \mp i\,\frac{m_0 c}{d_2} , 1        \right)\ .
\label{eq:3.21e}
\eea
\ese
These modes have the characteristics of a ferromagnetic magnon, with a quadratic dispersion relation and a
damping term that vanishes as $k^4$. 
In the limit of vanishing damping coefficients at fixed wave number the oscillation frequency remains the same, but
the damping changes. The eigenfrequencies in this limit are
\be
\tilde\Omega_{2,\pm} = \pm d_2 k^2 + \frac{i}{2}\left(\Gamma_0 a k^2 + \lambda_0 c k^4\right)\ .
\label{eq:3.22}
\ee
Note the damping term proportional to $k^2$, which is absent in the limit of asymptotically small wave number.

The previous results for the relaxational modes, Eqs.~(\ref{eqs:3.12}) through (\ref{eqs:3.14}), remain valid if
one puts $h=0$ at fixed $m_0$.

It is not obvious how the results given above cross over from the ones in Sec.~\ref{subsubsec:III.A.2} in the limit
$h\to 0$ at fixed $m_0$. In Appendix \ref{app:A} we give a solution of the eigenvalue problem represented by
Eq.~(\ref{eq:3.8}) that interpolates between the two results. 

\subsubsection{Collinear order parameters}
\label{subsubsec:III.B.2}

Of the two pairs of gapped modes shown in Eqs.~({\ref{eqs:3.17}), one remains gapped in the limit $h\to 0$ at
fixed $m_0$. However, $m_0$ should no longer be considered small. To zeroth order in the wave number we
find, instead of Eq.~(\ref{eq:3.17a}),
\bse
\label{eqs:3.23}
\be
\Omega_{1,\pm} = \pm w_2 m_0(n_0^2 - m_0^2) - i w_2 \Gamma_0 m_0^2 + O(k^2)\ .
\label{eq:3.23a}
\ee
For the other eigenfrequency, Eq.~(\ref{eq:3.17b}), both the propagating part and the damping part vanish
to $O(k)$ and $O(k^2)$, respectively, and we need to go to quartic order in $k$. An elementary but tedious
calculation yields
\bea
\Omega_{2,\pm} &=& \pm \left(\frac{n_0^2 a + m_0^2 c}{m_0}\,k^2  + \frac{n_0^2 a^2}{w_2 m_0^3}\,k^4\right) 
\nonumber\\ 
&&\hskip 20pt + i\lambda_0\,\frac{n_0^2 a + m_0^2 c}{m_0^2}\,k^4 + O(k^6)\ .\qquad
\label{eq:3.23b}
\eea
\ese
To lowest order in the wave number the corresponding eigenvectors are still given by Eqs.~(\ref{eq:3.17d}) - (\ref{eq:3.17g}).
We see that the gapped modes are fluctuations of the staggered magnetization that are gapped due to the coupling
to the collinear homogeneous magnetization. The gapless mode is a linear combination of staggered and homogeneous
fluctuations that are locked together and behave like a ferromagnetic magnon. 

The modes shown in Eqs.~(\ref{eqs:3.18}) and (\ref{eqs:3.19}) allow for a straightforward limit $h\to 0$ at fixed $m_0$
to be taken. We thus again obtain a diffusive mode with the diffusion constant given by Eq.~(\ref{eq:3.18b}) with $h=0$
and the eigenvectors given by Eqs.~(\ref{eq:3.18c}, \ref{eq:3.18d}), and a relaxation mode that is still given by
Eqs.~(\ref{eqs:3.19}).

\subsection{Non-conserved homogeneous magnetization: AFM order only}
\label{subsec:III.C}

We now discuss the case of a non-conserved homogeneous magnetization, which means that Eq.~(\ref{eq:2.24b}) gets
replaced by Eq.~(2.24b'). Although formally this amounts to replacing $\lambda_0 k^2$ by $\mu_0$, the result can in
in general not be obtained by performing this substitution in the results of Sec.~\ref{subsec:III.A}, because
the reality properties of the eigenvalue problems may change. As a result the problem gets quite involved, with many
different cases depending on parameter values. Since we are mainly interested in soft modes, we will derive and
discuss only those in detail. For relaxational modes, and for gapped propagating modes, we will list the eigenfrequencies
at zero wave number, but we will not discuss their $k$-dependence or the corresponding eigenvectors.

\subsubsection{Zero field}
\label{subsubsec:III.C.1}

In the absence of an external field, the kinetic equations for the transverse fluctuations now are, instead of Eq.~(\ref{eq:3.1}),
\be
\begin{pmatrix} i\Omega + \Gamma_0 a k^2  &  -n_0(t + w_1 n_0^2 + c k^2)  \\
                         n_0 a k^2                               &  i\Omega + \mu_0(t + w_1 n_0^2 + c k^2)  \\
                         \end{pmatrix}
\begin{pmatrix} \theta_{1,2} \\
                         \pi_{1,2}     \\
                         \end{pmatrix} = 0 \ .
\label{eq:3.24}
\ee  
For asymptotically small $k$ this yields two identical diffusive modes with eigenfrequencies
\bse
\label{eqs:3.25}
\be
\Omega_1 = \Omega_2 = i(\Gamma_0 + n_0^2/\mu_0)a\, k^2 + O(k^4)\ .
\label{eq:3.25a}
\ee
The corresponding right and left eigenvectors are
\bea
{\bm\nu}_{\text{R}1} &\equiv& \begin{pmatrix} \theta_1 \\ \pi_1 \\ \end{pmatrix}_{\text{R}1} = \begin{pmatrix} 1 \\ - n_0 a\,k^2/\mu_0(t + w_1 n_0^2) + O(k^4) \\ \end{pmatrix}\ ,
\nonumber\\
\label{eq:3.25b}\\
{\bm\nu}_{\text{R}2} &\equiv& \begin{pmatrix} \theta_2 \\ \pi_2 \\ \end{pmatrix}_{\text{R}2}= {\bm\nu}_{\text{R}1} 
\label{eq:3.25c}\\
{\bm\nu}_{\text{L}1} &\equiv& \left(\theta_1,\pi_1\right)_{\text{L}1} = \left(\mu_0/n_0 + O(k^2),1\right)\ ,
\label{eq:3.25d}\\
{\bm\nu}_{\text{L}2} &\equiv& \left(\theta_1,\pi_1\right)_{\text{L}2} = {\bm\nu}_{\text{L}1}\ .
\label{eq:3.25e}
\eea
\ese
We note in passing that for larger wavenumbers Eq.~(\ref{eq:3.24}) describes two propagating modes, see
the discussion in Sec.~\ref{sec:IV}.
The other solution of the quadratic equation yields two identical relaxational modes with eigenfrequencies
\be
\Omega_3 = \Omega_4 = i\mu_0(t + w_1 n_0^2) + O(k^2)\ ,
\label{eq:3.26}
\ee
In addition, the analogs of Eqs.~(\ref{eq:3.2}) and (\ref{eq:3.3}) yield two more relaxational modes. One is a pure 
$\theta_3$ mode with eigenfrequency
\be
\Omega_5 = i\Gamma_0(2 u n_0^2 + a\,k^2)\ ,
\label{eq:3.27}
\ee
and one is a pure $\pi_3$ mode with eigenfrequency
\be
\Omega_6 = i\mu_0\left[t + (w_1+w_2)n_0^2 + c\,k^2\right]\ .
\label{eq:3.28}
\ee
Note that the mode spectrum is qualitatively different compared to the case of a conserved homogeneous magnetization:
There are no propagating spin waves; instead, the transverse fluctuations form one diffusive mode and one relaxational one.

\subsubsection{Transverse field}
\label{subsubsec;III.C.2}

The relevant kinetic equations for this case are obtained by replacing $\lambda_0 k^2$ in Eqs.~(\ref{eq:3.7}, \ref{eq:3.8}) by $\mu_0$.
The first $3\times 3$ system yields one diffusive mode with eigenfrequency
\begin{widetext}
\bse
\label{eqs:3.29}
\be
\Omega_1 = i\,D_1 k^2 + O(k^4)
\label{eq:3.29a}
\ee
where the diffusion constant is given by
\be
D_1 = a\! \left[\Gamma_0 + \frac{m_0^2}{\Gamma_0}\,(1 - w_1/u) + \frac{n_0^2}{\mu_0}\,\frac{h + 2 m_0^3(v-w_1^2/u)}{h + 2 m_0^3 v}\right]\!.
\label{eq:3.29b}
\ee
The corresponding right and left eigenvectors are
\bea
{\bm\nu}_{\text{R}1} &=& \begin{pmatrix} \theta_1 \\
                                                   \theta_3 \\
                                                   \pi_1 \\
                                                   \end{pmatrix}_{\text{R}1}                                                 
                      = \begin{pmatrix} 1 \\
                                                  \left(\frac{1}{2 u \Gamma_0 n_0^2} + \frac{w_1/u}{\mu_0(h/m_0 + 2 v m_0^2)}\right) m_0 a k^2 + O(k^4) \\
                                                  \frac{-n_0 a k^2}{\mu_0(h/m_0 + 2 v m_0^2)} + O(k^4)\\
                                                  \end{pmatrix}  ,                                             
\label{eq:3.29c}\\
{\bm\nu}_{\text{L}1} &=& \left(\theta_1,\theta_3,\pi_1\right)_{\text{\text{L}}1} = \left(\frac{\mu_0 (h/m_0 + 2 m_0^2 v)}{n_0 (h/m_0 + 2 m_0^2(v - w_1^2/u))} + O(k^2), 
                            \frac{-\mu_0 m_0 (h/m_0 + 2 m_0^2 v)(1-w_1/u)}{\Gamma_0 n_0 (h/m_0 + 2 m_0^2(v - w_1^2/u))} + O(k^2), 1\right).
\nonumber\\                            
\label{eq:3.29d}
\eea
\ese
\end{widetext}
In addition, we find two relaxational modes  with eigenfrequencies
\be
\Omega_2 = i 2 \Gamma_0 u n_0^2 + O(k^2)\ ,
\label{eq:3.30}
\ee
and
\be
\Omega_3 =  i \mu_0(h/m_0 + 2 m_0^2 v) + O(k^2)\ ,
\label{eq:3.31}
\ee

Now consider the $3\times 3$ system that is the analog of Eq.~(\ref{eq:3.8}). Here the energy scale
$w_2 \mu_0 n_0^2$ competes with $h$, and for small $h$ there are neither propagating nor diffusive modes. 
Instead we find three relaxational modes with eigenfrequencies
\bea
\Omega_4 &=& i \mu_0 h/m_0 + O(\epsilon^2)\ ,
\label{eq:3.32}\\
\Omega_5 &=& i \mu_0(h/m_0 + w_2 n_0^2) + O(\epsilon^2)\ ,
\label{eq:3.33}\\
\Omega_6 &=& i w_2 h m_0 \frac{n_0^2 + \Gamma_0 \mu_0}{\mu_0(h/m_0 + w_2 n_0^2)} + O(\epsilon^4)\ ,
\label{eq:3.34}
\eea
with $\epsilon = O(h,m_0).$

\subsubsection{Longitudinal field}
\label{subsubsec:III.C.3}

The $2\times 2$ matrix that is analogous to Eq.~(\ref{eq:3.15}) yields two relaxational modes. To zeroth order in $h$ and $m_0$
the eigenfrequencies are
\be
\Omega_1 = i 2\Gamma_0 u n_0^2 + O(\epsilon^2)
\label{eq:3.35}
\ee
and
\be
\Omega_2 = i \mu_0(h/m_0) + O(\epsilon^2)\ .
\label{eq:3.36}
\ee

From the $4\times 4$ matrix that is the analog of Eq.~(\ref{eq:3.16a}) one finds four additional relaxational modes. 
To lowest order in $\epsilon = O(h,m_0)$ the corresponding eigenfrequencies are
\bea
\Omega_3 &=& \Omega_4 = i\mu_0 (h/m_0 - w_2 n_0^2) + O(\epsilon)\ ,
\label{eq:3.37}\\
\Omega_5 &=& \Omega_6 = -i w_2 h m_0\,\frac{n_0^2 + \Gamma_0 \mu_0}{\mu_0(h/m_0 - w_2 n_0^2)} + O(\epsilon^3)\ .
\nonumber\\
\label{eq:3.38}
\eea

\subsection{Non-conserved homogeneous magnetization: Coexisting AFM and homogeneous orders}
\label{subsec:III.D}

\subsubsection{Orthogonal order parameters}
\label{subsubsec:III.D.1}

The $3\times 3$ matrix that is analogous to Eq.~(\ref{eq:3.7}) yields one diffusive mode:
\bse
\label{eqs:3.39}
\be
\Omega_1 = i \left(\Gamma_0 + n_0^2/\mu_0 + m_0^2/\Gamma_0\right) a\,k^2 + O(k^4)\ .
\label{eq:3.39a}
\ee
The corresponding right and left eigenvectors are
\be
{\bm\nu}_{{\rm R}1} \equiv \begin{pmatrix} \theta_1 \\
                                                                    \theta_3 \\
                                                                    \pi_1      \\
                                                                           \end{pmatrix}_{\text{R}1}
                       = \begin{pmatrix} 1 \\
                                                    \frac{w_1 \Gamma_0 n_0^2 +v \mu_0 m_0^2}{2\Gamma_0\mu_0 n_0^2 m_0(uv - w_1^2)}\,a k^2  + O(k^4)\\
                                                    \frac{-(u \Gamma_0 n_0^2 + w_1 \mu_0 m_0^2)} {2\Gamma_0\mu_0 n_0 m_0^2(uv - w_1^2)} \,a k^2+ O(k^4) \\
                                                      \end{pmatrix}\ ,
\label{eq:3.39b}
\ee
\bea
{\bm\nu}_{\text{L}1} &=& \left(\theta_1,\pi_2,\pi_3\right)_{\text{L}1} 
\nonumber\\
&=& \left(\frac{\mu_0}{n_0} + O(k^2),-\frac{\mu_0 m_0}{\Gamma_0 n_0} + O(k^2),1\right). \quad
\label{eq:3.39c}
\eea
\ese
Note that if we put $\mu_0 = \lambda_0 k^2$ we can {\em not} relate these expressions to Eqs.~(\ref{eqs:3.9}, \ref{eqs:3.20}), since the nature
of the mode has changed.

The other two modes are relaxational with eigenfrequencies
\bea
\Omega_{2,3} &=& i\biggl(u \Gamma_0 n_0^2 + v \mu_0 m_0^2 
\nonumber\\
&& \left. \mp \sqrt{(u \Gamma_0 n_0^2 - v\mu_0 m_0^2)^2 + 4 w_1^2 \Gamma_0 \mu_0 n_0^2}\right).
\nonumber\\
\label{eq:3.40}
\eea
From the $3\times 3$ matrix that is analogous to Eq.~(\ref{eq:3.8}) we obtain two soft modes. Depending on parameter values,
they can be gapless propagating with both the oscillation frequency and the damping coefficient proportional to $k^2$:
\begin{widetext}
\bse
\label{eqs:3.41}
\be
\Omega_{4,\pm} = \frac{\pm k^2}{2(\Gamma_0 m_0^2 + \mu_0 n_0^2)}\,\sqrt{4 d_2^2(\Gamma_0 m_0^2 + \mu_0 n_0^2)^2 - \mu_0^2(\Gamma_0 a - \mu_0 c)^2 n_0^4} 
                              + i\mu_0\,\frac{\Gamma_0(n_0^2 a + 2 m_0^2 c) + \mu_0 n_0^2 c}{2(\Gamma_0 m_0^2 + \mu_0 n_0^2)}\,k^2     + O(k^4)\ ,
\label{eq:3.41a}
\ee
with $d_2$ from Eq.~(\ref{eq:3.21b}) and right and left eigenvectors
\be
{\bm\nu}_{{\rm R}4,\pm} \equiv \begin{pmatrix} \theta_2 \\
                                                                    \pi_2 \\
                                                                    \pi_3      \\
                                                                           \end{pmatrix}_{\text{R}4,\pm}
                       = \begin{pmatrix} 1 \\
                                                  \pm i \sqrt{m_0^2/n_0^2 + \mu_0/\Gamma_0} + O(k^2)  \\
                                                     m_0/n_0 + O(k^2) \\
                                                      \end{pmatrix}\ ,
\label{eq:3.41b}
\ee
\be
{\bm\nu}_{{\rm L}4,\pm} \equiv \left(\theta_2,\pi_2,\pi_3\right)_{\text{L}4,\pm} = \left(\frac{n_0 \mu_0}{m_0 \Gamma_0} + O(k^2), \mp i \sqrt{1 + \mu_0 n_0^2/\Gamma_0 m_0^2} + O(k^2),1\right)\ .
\label{eq:3.41c}
\ee
\ese
For $\mu_0 \to \lambda_0 k^2$ this is consistent with Eqs.~(\ref{eqs:3.21}). However there is no guarantee that the expression under the
square root in Eq.~(\ref{eq:3.41a}) is positive. If it is not, then one has instead two diffusive modes,
\bse
\label{eqs:3.42}
\be
\tilde\Omega_{4,\pm} = i D_{4,\pm} k^2 + O(k^4)
\label{eq:3.42a}
\ee
with diffusion coefficients
\be
D_{4,\pm} = \frac{1}{2(\Gamma_0 m_0^2 + \mu_0 n_0^2)}\,\left[\Gamma_0 \mu_0(n_0^2 a + 2 m_0^2 c) + \mu_0^2 n_0^2 c 
                     \pm \sqrt{\mu_0^2(\Gamma_0 a - \mu_0 c)^2 n_0^4 - 4 d_2^2(\Gamma_0 m_0^2 + \mu_0 n_0^2)^2 } \right]\ .
\label{eq:3.42b}
\ee      
\ese               
In addition, there is one relaxational mode with eigenfrequency
\be
\Omega_5 = i w_2 (\Gamma_0 m_0^2 + \mu_0 n_0^2)\ .
\label{eq:3.43}
\ee

\subsubsection{Collinear order parameters}
\label{subsubsec:III.D.2}

The $4\times 4$ problem that is analogous to Eq.~(\ref{eq:3.16a}) yields a pair of gapless propagating modes with eigenfrequencies
\bse
\label{eqs:3.44}
\bea
\Omega_{1,\pm} &=& \pm m_0(m_0^2 c + n_0^2 a)\frac{(m_0^2-n_0^2)^2+ \Gamma_0^2 m_0^2 + \mu_0^2 n_0^2 + 2\Gamma_0\mu_0 n_0^2}
                                                                                    {m_0^2(m_0^2 - n_0^2)^2 + (\Gamma_0 m_0^2 + \mu_0 n_0^2)^2}\,k^2 
\nonumber\\                                                                                
&&                               + i\mu_0(m_0^2 c + n_0^2 a)\,\frac{(m_0^2 - n_0^2)^2 + \Gamma_0^2 m_0^2 + \Gamma_0 \mu_0 n_0^2}
                                                                                       {m_0^2(m_0^2 - n_0^2)^2 + (\Gamma_0 m_0^2 + \mu_0 n_0^2)^2}\,k^2 + O(k^4)\ .
\label{eq:3.44a}
\eea
The right and left eigenvectors are, at $k=0$,\cite{eigenvectors_footnote}
\bea
{\bm\nu}_{{\rm R}1,\pm} &\equiv& \begin{pmatrix} \theta_1 \\
                                                                           \theta_2 \\
                                                                           \pi_1  \\
                                                                           \pi_2 \\
                                                                           \end{pmatrix}_{\rm{R}1,\pm}   
                                             = \begin{pmatrix} 1          \\
                                                                          \mp i   \\
                                                                          \pm i\, m_0/n_0  \\
                                                                          m_0/n_0  \\
                                                  \end{pmatrix}\ ,
\label{eq:3.44b} \\                                                                           
{\bm\nu}_{{\rm L}1,\pm} &\equiv&  \left(\theta_1, \theta_2, \pi_1, \pi_2\right)_{{\rm L}1,\pm} 
\nonumber\\
                                      &=&  \left(\frac{\Gamma_0 \mu_0 n_0 m_0 \mp i\mu_0 n_0(n_0^2 - m_0^2)}{(n_0^2 - m_0^2)^2+m_0^2 \Gamma_0^2},
                                                      \frac{\mu_0 n_0(n_0^2 - m_0^2) \pm i \Gamma_0 \mu_0 n_0 m_0}{(n_0^2 - m_0^2)^2+m_0^2 \Gamma_0^2}, \mp i, 1 \right)\ .
\label{eq:3.44c}
\eea
\ese
In addition, there is a pair of gapped propagating modes with eigenfrequencies
\be
\Omega_{2,\pm} = \pm w_2 m_0(m_0^2 - n_0^2) - i w_2 (\Gamma_0 m_0^2 + \mu_0 n_0^2) + O(k^2)\ .
\label{eq:3.45}
\ee
Note that the damping coefficient does not vanish at $k=0$.

The $2\times 2$ problem that is analogous to Eq.~(\ref{eq:3.15}) now yields two relaxational modes with eigenfrequencies
\be
\Omega_{3,4} = i\left( u\Gamma_0 n_0^2 + v\mu_0 m_0^2  \pm \sqrt{(u\Gamma_0 n_0^2 - v\mu_0 m_0^2)^2 + 4(w_1+w_2)^2\Gamma_0\mu_0 n_0^2 m_0^2}\right)\ .
\label{eq:3.46}
\ee
\end{widetext}

\section{Summary, and Discussion}
\label{sec:IV}

In summary, we have used time-dependent Ginzburg-Landau theory, with the basic equation of motion for magnetic
moments the only input, to determine the long-wavelength spin dynamics of antiferromagnets in various physical situations. We have
considered purely antiferromagnetic order subject to an external magnetic field, and have distinguished between the
cases of a conserved and a non-conserved homogeneous magnetization. We have also considered the case of
coexisting antiferromagnetic and ferromagnetic orders, as it occurs, for example, in ferrimagnets and canted magnets. Our results
are summarized in Tables~\ref{tab:1} and \ref{tab:2}.
\begin{table}[h]
\caption{Number of modes of various types for antiferromagnetic order with the homogeneous magnetization conserved/non-conserved.}
\begin{ruledtabular}
\begin{tabular}{c|cccc}
Field & \multicolumn{4}{c}{Modes}
\\
\hline
\\
                   & \multicolumn{2}{c}{propagating} & diffusive & relaxational
\\
                   & gapless & gapped                      &               &
\\
\cline{2-5}
\\
zero            & 4\,/\,0     & 0\,/\,0                        & 1\,/\,2      & 1$^*$/\,4
\\
transverse  & 2\,/\,0     & 2\,/\,0                        & 0\,/\,1      & 2\,/\,5
\\
longitudinal & 0\,/\,0     & 4\,/\,0                        & 1\,/\,0     & 1\,/\,6
\\
\hline\hline\\
& \multicolumn{4}{l} {$^*$ See the comments after Eq.~(\ref{eq:3.6}).}
\end{tabular}
\end{ruledtabular}   
\label{tab:1}
\end{table}       
\begin{table}[h]
\caption{Number of modes of various types for coexisting antiferromagnetic and ferromagnetic orders with the homogeneous magnetization conserved/non-conserved.}
\begin{ruledtabular}
\begin{tabular}{c|cccc}
Order\\Parameters & \multicolumn{4}{c}{Modes}
\\
\hline
\\
                   & \multicolumn{2}{c}{propagating} & diffusive & relaxational
\\
                   & gapless & gapped                      &               &
\\
\cline{2-5}
\\
orthogonal  & 4\,/\,2\ \text{or}\ 0     & 0\,/\,0                        & 0\,/\,1\ \text{or}\ 3      & 2\,/\,3
\\
collinear & \hskip -19pt 2\,/\,2     & 2\,/\,2                        & \hskip -19pt 1\,/\,0     & 1\,/\,2
\\
\end{tabular}
\end{ruledtabular}   
\label{tab:2}
\end{table}       

\subsection{Nature of spin excitations}
\label{subsec:IV.A}

One striking aspect of these results is the qualitative difference between the cases of a conserved and a non-conserved
homogeneous magnetization, respectively. In the former case in a zero external field the transverse 
fluctuations form two pairs of gapless propagating spin waves with a linear dispersion relation. The longitudinal order-parameter
fluctuations are relaxational, whereas the longitudinal fluctuations of the homogeneous magnetization are diffusive. In a transverse
field one of the spin-wave pairs remains gapless with a linear dispersion, the other one acquires a gap that is proportional to the
field, and there is no diffusive mode. In a longitudinal field one finds two pairs of gapped propagating spin waves and one
diffusive mode. In all cases there thus are two pairs of propagating spin waves. If the homogeneous magnetization is not
conserved, in contrast, there are no propagating spin waves and the only soft modes are two diffusive modes in zero field
and one diffusive mode in the transverse-field
case. (To avoid misunderstandings we reiterate that there is no experimental control over the orientation of the field with
respect to the order parameter; the sign of the Landau parameter $w_2$ determines which case is realized in any given
system.) This case is realized, for instance, in an antiferromagnet with magnetic impurities. The results of 
Ref.~\onlinecite{Brenig_Kampf_1991} obtained by applying linear spin-wave theory to this case, which found propagating 
modes, are thus not valid in the long-wavelength limit. Propagating spin waves are reconstituted, however, 
above a threshold wave number, as we discuss next. 

An interesting feature of the antiferromagnet in zero field is the crossover between the cases of a conserved
and a non-conserved homogeneous magnetization in the limit $\mu_0\to 0$ or, for fixed small $\mu_0$, with
increasing wave number. Consider the eigenproblem posed
by Eq.~(\ref{eq:3.24}), and assume that the damping is small in the sense that $\mu_0 \Gamma_0 \ll n_0^2$ and
$\mu_0^2 \ll n_0^2 a/c$. Then the reality properties of the quadratic equation change at a critical value of the
wave number $k$, and the diffusive modes become propagating for $k > k_{\rm{c}}$ where $k_{\rm{c}} = \mu_0\sqrt{(t+w_1 n_0^2)/a n_0^2}$.
In more physical terms, the threshold wave number is $k_{\rm{c}}/k_0 = 1/E_0\tau_{\rm{mag}}$, with $k_0$ and $E_0$
microscopic wave-number and energy scales, respectively, and $1/\tau_{\rm{mag}} \propto \mu_0$ the relaxation rate associated
with the magnetic impurities. Similarly, the two relaxational modes from Eq.~(\ref{eq:3.26}) become propagating, and
collectively these four modes cross over to the four propagating spin waves that characterize the antiferromagnet
with a conserved homogeneous magnetization, Eq.~(\ref{eq:3.4a}). For metals at low temperatures we expect $k_0$
and $E_0$ to be on the order of the Fermi wave number and the Fermi energy, respectively, and the value of $k_{\rm{c}}$
can be quite small.

Alternatively, let $\mu_0\to 0$ in the limit of asymptotically small $k$. Then the two diffusive modes and two of the 
relaxational modes listed in Table~\ref{tab:1} for the zero-field non-conserved case turn into the four propagating
modes of the conserved case, one of the relaxational modes becomes diffusive, and the remaining mode remains
relaxational.

\subsection{Properties of propagating spin waves}
\label{subsec:IV.B}

\subsubsection{Models, and their restrictions}
\label{subsubsec:IV.B.1}

The long-wavelength and low-frequency properties of the Heisenberg model for an antiferromagnetic nearest-neighbor coupling are captured
by an effective field theory that takes the form of a nonlinear sigma model.\cite{Chakravarty_Halperin_Nelson_1989,
Fisher_1989, Sachdev_1999} It needs to be noted that this model {\em always} has the staggered magnetization
point in a direction perpendicular to an external magnetic field, i.e., the simple Heisenberg model captures only the case
$w_2>0$ (the transverse-field case) in Eq.~(\ref{eq:2.4}) and is less general than the Landau theory we have used. 
The sigma model correctly describes the two gapless and two gapped
modes shown in Table~\ref{tab:1}.\cite{Sachdev_1999} In Ref.~\onlinecite{Hasselmann_et_al_2007} it was pointed out that
the field-dependence of the dispersion relations predicted by the sigma model does not agree with the results of
spin-wave theory, even though the qualitative features of the spin waves are correct. A comparison with the coefficients
$c_1$ and $c_2$ in Eqs.~(\ref{eq:3.9d}) and (\ref{eq:3.11d}) shows that their field dependence as obtained from the
hydrodynamic equations is still substantially more complicated than the results of the spin-wave theory employed in
Ref.~\onlinecite{Hasselmann_et_al_2007}. In the sigma model, both of these coefficients are replaced by their values at $h=0$.

We  have truncated the Landau free energy, Eq.~(\ref{eq:2.4}), at biquadratic order. The property that the
staggered magnetization is either parallel or perpendicular to the homogeneous magnetization is a consequence of this
truncation. For instance, keeping a term proportional to $({\bm n}\cdot{\bm m})^4$ allows for states where ${\bm n}$ and
${\bm m}$ are neither collinear nor orthogonal to each other. This can become important in large magnetic fields, when
${\bm m}$ is no longer small, see, e.g., Ref.~\onlinecite{Alicea_Chubukov_Starykh_2009}. If this is important for a
specific purpose one can keep higher order terms in the free-energy functional and repeat the analysis of the
hydrodynamic equations, which is completely general.

We also note that our discussion applies to systems in spatial dimensions $d\geq 2$. Antiferromagnetic spin chains
show qualitatively different behavior that requires a special treatment; see, e.g., Ref.~\onlinecite{Oshikawa_Yamanaka_Affleck_1997}.

\subsubsection{Damping}
\label{subsubsec:IV.B.2}

In Eqs.~(\ref{eqs:2.24}) we have used the standard Landau-Lifshitz form for the damping terms.\cite{Landau_Lifshitz_1935, 
Hohenberg_Halperin_1977, Ma_1976} Landau and Lifshitz considered a $\phi^4$-theory and enforced a time-independent
modulus of the magnetic moment by writing Eq.~(\ref{eq:2.21}) with the damping term added as
\bea
\partial_t{\bm M} &=& {\bm M}\times\frac{\delta F}{\delta{\bm M}} - \Gamma\left[\frac{\delta F}{\delta{\bm M}} - {\bm M}\left({\bm M}\cdot\frac{\delta F}{\delta{\bm M}}\right)/{\bm M}^2\right]
\nonumber\\
&=& {\bm M}\times\frac{\delta F}{\delta{\bm M}} + \Gamma {\bm M}\times\left({\bm M}\times\frac{\delta F}{\delta{\bm M}}\right)/{\bm M}^2\ .
\label{eq:4.1}
\eea
Gilbert later proposed to replace the ${\bm M}\times(\delta F/\delta{\bm M})$ in the damping term by $\partial_t{\bm M}$. The
resulting Landau-Lifshitz-Gilbert equation,
\be
\partial_t{\bm M} = {\bm M}\times\frac{\delta F}{\delta{\bm M}} + \Gamma {\bm M}\times\partial_t{\bm M}/{\bm M}^2\ ,
\label{eq:4.2}
\ee
is very popular on phenomenological grounds, but its consistency with basic principles of irreversible thermodynamics
is questionable, see, e.g., Ref.~\onlinecite{Saslow_2009}. Equation (\ref{eq:4.2}) can be mapped onto Eq.~(\ref{eq:4.1})
at the expense of making the prefactor of the Bloch term depending on the damping coefficient.\cite{Aharoni_2000} This
observation underscores the fact that the Gilbert modification does not have the standard hydrodynamic form, but it also
means that, as far as the nature of
the spin excitations is concerned, the difference between Eqs.~(\ref{eq:4.1}) and (\ref{eq:4.2}) is irrelevant.

Regarding the nature of the damping, in a simple antiferromagnet in zero field the damping is quadratic in the wavenumber,
and thus always small, in the long-wavelength limit, compared to the oscillation frequency, which is linear in $k$, see Eq.~(\ref{eq:3.4a}).
This is important for ensuring the correct relation between the static Goldstone modes and the time-correlation functions,
as we demonstrate in Appendix~\ref{app:B}. In a simple ferromagnet, the damping is proportional to $k^2$, and the damping to $k^4$. 
For an antiferromagnet with a conserved homogeneous magnetization in a transverse field, the gapless propagating modes still have a linear dispersion relation and both
they and the gapped modes have a quadratic damping, see Eqs.~(\ref{eqs:3.9}, \ref{eqs:3.11}), but in a longitudinal field
one of the pairs of gapped modes has a damping coefficient that is nozero at $k=0$, see Eq.~(\ref{eq:3.17a}). For coexisting
antiferromagnetic and homogeneous order there are ferromagnon-like spin waves, with a quadratic oscillation frequency and
quartic damping, see Eqs.~(\ref{eq:3.21a}) and (\ref{eq:3.23b}).

In the context of damping it is also interesting to see how the damping term that appears in effective field theories for metals,\cite{Hertz_1976}
which is often referred to as Landau damping in analogy to the corresponding effect in a collisionless classical plasma, is related 
to the hydrodynamic equations. In the case of a ferromagnetic metal with nonmagnetic impurities the Landau-damping term in the paramagnon
propagator has the form $\vert\Omega\vert/D k^2$, where $D$ is the diffusion coefficient related to the diffusive dynamics of
the conduction electrons in the spin-triplet channel. This corresponds to the damping term in the hydrodynamic equations
for the case of a conserved order parameter, see Eq.~(\ref{eq:2.24b}). For an antiferromagnetic metal, or for a ferromagnet
with magnetic impurities, the corresponding term is $\vert\Omega\vert\tau$, with $\tau$ a $k$-independent relaxation time. 
This corresponds to the damping term for a non-conserved order parameter, see Eqs.~(\ref{eq:2.24a}) and (2.24b'). 

In a clean metallic ferromagnet the Landau-damping term has the form $\vert\Omega\vert/\vF k$, with $\vF$ the Fermi velocity.\cite{Hertz_1976}
In order to see how this case fits into the hydrodynamic description, we note that Eq.~(\ref{eq:2.24b}) implies a long-wavelength
susceptibility of the form
\be
\chi(k,i\Omega) \propto -i\Omega/\lambda_0 k^2 + k^2\ ,
\label{eq:4.3}
\ee
for the conserved and non-conserved cases, respectively. In our case $\chi$ is the spin susceptibility, but the following
discussion holds more generally for any order-parameter susceptibility. In a clean metallic system at $T=0$ the kinetic 
coefficient $\lambda_0$ does not exist in the limit of zero frequency and wave number and scales as 
$\lambda_0 \sim 1/(\Omega + k^{z_{\lambda}})$, with $z_{\lambda}$ the dynamical exponent characteristic of the
kinetic coefficient. In the conserved case, $z_{\lambda} \geq 2$ leads to $\Omega \sim k^2$, and $z_{\lambda} < 2$ 
leads to $\Omega \sim k^{4-z_{\lambda}}$. For the conduction electrons in a metal one has $\Omega \sim k$, and as 
long as the order parameter couples to the conduction electrons one therefore expects $z_{\lambda} = 1$. Effectively, 
we thus have $\lambda_0 \sim 1/k$,\cite{Belitz_Kirkpatrick_Rosch_2006a} which is consistent with the above form of 
the Landau-damping term.

\subsection{Outlook: Nonlinear effects, and consequences for quantum phase transitions}
\label{subsec:IV.C}

The current paper lays the groundwork for several investigations of properties of quantum antiferromagnets. For instance, 
Ref.~\onlinecite{Bharadwaj_Belitz_Kirkpatrick_2016} considered the coupling of the spin waves to the longitudinal fluctuations,
and the resulting behavior of the longitudinal susceptibility and the dynamical structure factor. Within the current formalism,
these effects are due to the nonlinearities in the hydrodynamic equations that we have neglected (see also the remark after
Eq.~(\ref{eq:3.6})). Due to limitations inherent in the nonlinear sigma model used in Ref.~\onlinecite{Bharadwaj_Belitz_Kirkpatrick_2016} 
the only cases considered were those of zero field, and antiferromagnetic order in a transverse field. An analogous investigation
of all cases discussed in the present paper would be of interest, especially for the dynamical structure factor, which is directly
measurable by neutron scattering. 

A renormalized mean-field theory for quantum ferromagnets predicted that the quantum phase transition from a paramagnet to
a ferromagnet or ferrimagnet in clean metals is necessarily first order.\cite{Belitz_Kirkpatrick_Vojta_1999, Kirkpatrick_Belitz_2012b}
This prediction has been confirmed by experiments on many different materials.\cite{Brando_et_al_2016a} The present paper
makes possible analogous theories for other quantum phase transitions. For instance, one expects the prediction of a universal
tricritical point\cite{Kirkpatrick_Belitz_2012b} to hold for canted magnets (the case $w_2>0$ in our notation) in addition to
ferrimagnets ($w_2>0$). Furthermore, it allows for a treatment of the quantum phase transition from a metallic ferromagnet
to an antiferromagnet, of which there are various known examples.\cite{Brando_et_al_2016a}

Another class of phase transitions that is of interest in this context is the spin-flop transition in uniaxial antiferromagnets as a function
of an applied external field. The history of these classical transitions goes back to Ne{\'e}l in the 1930s; they have been studied extensively
from the viewpoint of classical phase-transition theory\cite{Fisher_Nelson_1974} and continue to be of great interest, see
Ref.~\onlinecite{Bogdanov_Zhuravlev_Rossler_2007} and references therein. They usually are first order, but can be second order
in certain materials.\cite{Yokosuk_et_al_2015} An investigation of the corresponding quantum phase transitions would be of interest.

\acknowledgments
This work was supported by the NSF under Grants No. DMR-1401410 and No. DMR-1401449. 
Part of this work was performed at the Aspen Center for Physics, which is supported by the NSF under Grant No. PHY-1066293. $\mbox{\hskip 100pt}$

\appendix

\section{Interpolating solution for the transverse-field and orthogonal-order-parameters cases}
\label{app:A}

As we saw in Secs.~\ref{subsubsec:III.A.2} and \ref{subsubsec:III.B.1}, the nature of the solution of Eq.~(\ref{eq:3.8}) changes
qualitatively if one considers the limit $h\to 0$ for fixed $m_0$. To see the crossover, one can of course solve the cubic equation
exactly, but this is not very illuminating. It is more useful to keep all terms that contribute to leading order in $k$ and $h$ in either
the pure AFM case or the coexisting-orders case. The eigenfrequencies then can be found by solving linear equations only. The
result is
\begin{widetext}
\bse
\label{eqs:A.1}
\bea
\Omega_{2,\pm} &=& \pm \sqrt{(h + m_0 c k^2)^2 + c_2^2 k^2 + n_0^2 a c k^4} 
\nonumber\\
&& + \frac{i}{2}\,k^2 \left[\lambda_0(h/m_0 + c k^2) + \frac{(h/m_0)(\lambda_0 h^2 + \Gamma_0 n_0^2 a^2 k^2) 
                + \Gamma_0^2 \lambda_0 w_2^2 m_0^2 (a n_0^2 + c m_0^2) k^2}{h^2 + c_2^2 + \Gamma_0^2 w_2^2 m_0^4}\right]\ .
\label{eq:A.1a}                
\eea
The corresponding right and left eigenvectors are
\be
{\bm\nu}_{R2,\pm} \equiv \begin{pmatrix} \theta_2  \\
                                                                  \pi_2       \\
                                                                  \pi_3 \\
                                                                  \end{pmatrix}_{R2,\pm} = \begin{pmatrix} 1 \\
                                                                         \pm i\,\frac{\sqrt{(h + m_0 c k^2)^2 + c_2^2 k^2 + n_0^2 a c k^4}}{n_0(h/m_0 + c k^2)} \\
                                                                         m_0/n_0 \\
                                                                         \end{pmatrix}\ .
\label{eq:A.1b}
\ee   
\be
{\bm\nu}_{L2,\pm} \equiv  \left(\theta_2,\pi_2,\pi_3\right)_{L2,\pm} = \left(\frac{n_0 a k^2}{h + m_0 c k^2}, \frac{\mp i}{h + m_0 c k^2}\,
        \sqrt{(h + m_0 c k^2)^2 + (h/m_0 + c k^2)n_0^2 a k^2}, 1\right)\ .
\label{eq:A.1c}
\ee
\ese              
\end{widetext}
These expressions are valid to leading order in $k$ and $h$ for both the AFM case, where the oscillation frequency is of $O(h,k)$ and the
damping is of $O(k^2)$, and the coexisting-orders case, where the oscillation frequency is of $O(k^2)$ and the damping is of $O(k^4)$.
They correctly interpolate between Eqs.~(\ref{eqs:3.11}) and (\ref{eqs:3.21}). To linear order in $\Gamma_0$ and $\lambda_0$ in the
coexisting-orders case one recovers Eq.~(\ref{eq:3.22}).

\section{Time correlation functions, and the fluctuation-dissipation theorem}
\label{app:B}

We mentioned at the end of Sec.~\ref{sec:II} that we have chosen to calculate response functions. It is illustrative to consider
the related problem of calculating time correlation functions. To this end we add Langevin forces ${\bm f}_n$ and ${\bm f}_m$ on 
the right-hand sides of Eq.~(\ref{eq:2.24a}) and (\ref{eq:2.24b}), respectively. These are random forces that are characterized by
Gaussian distributions with second moments
\bse
\label{eqs:B.1}
\bea
\langle f_n^i({\bm x},t)\,f_n^j({\bm x}',t') \rangle &=& 2T\,\Gamma_0\,\delta_{ij}\delta({\bm x}-{\bm x}')\,\delta(t - t')\ ,
\nonumber\\
\label{eq:B.1a}\\
\langle f_m^i({\bm x},t)\,f_m^j({\bm x}',t') \rangle &=& -2T\,\lambda_0 {\bm\nabla}^2\,\delta_{ij}\delta({\bm x}-{\bm x}')\,\delta(t - t'),
\nonumber\\
\label{eq:B.1b}\\
\langle f_n^i({\bm x},t)\,f_m^j({\bm x}',t') \rangle &=& 0\ ,
\label{eq:B.1c}
\eea
\ese
where $T$ is the temperature. For simplicity, we consider only classical systems in this appendix; for a discussion of a quantum
Langevin equation, see Ref.~\onlinecite{Ford_Lewis_O'Connell_1988}. These relations guarantee
the validity of the fluctuation-dissipation theorem.

We now illustrate the use of
this formalism for the simple case of an AFM in zero field, Sec.~\ref{subsubsec:III.A.1}; the other cases can be analyzed analogously.

Consider Eq.~(\ref{eq:3.1}) and add the fluctuating forces. Then we have, structurally,
\be
M\,{\bm\psi} = {\bm f}\ ,
\label{eq:B.2}
\ee
where $M$ denotes the $2\times 2$ matrix, ${\bm\psi}$ comprises the fluctuations $\theta$ and $\pi$, and $\bm f$ the appropriate
components of the fluctuating forces ${\bm f}_n$ and ${\bm f}_m$. 
Multiplying with the left eigenvector ${\bm\nu}_{{\rm L}1,\pm}$, Eq.~(\ref{eq:3.4e}), yields
\bse
\label{eqs:B.3}
\be
\Lambda_{1\pm}\, \psi_{1\pm} = \pm i(n_0a/c_1) k f_n^1 + f_m^1\ ,
\label{eq:B.3a}
\ee
where
\be
\psi_{1\pm} = \pm i(n_0 a/c_1) k\, \theta_1 + \pi_1\
\label{eq:B.3b}
\ee
and 
\be
\Lambda_{1\pm} = i\Omega \pm i c_1 k + \frac{1}{2}\Gamma_1 a k^2
\label{eq:B.3c}
\ee
\ese
is the eigenvalue of the matrix $M$ that corresponds to the eigenvector ${\bm\nu}_{\rm{L},1\pm}$, see Eqs.~(\ref{eqs:3.4}).
For the correlation function of $\psi_{1\pm}$ we thus have
\be
\langle \psi_{1\pm}\,\psi_{1\pm}^*\rangle_{\Omega,{\bm k}} =  \,\frac{-2T(n_0 a/c_1)^2 \Gamma_1 k^2}{(\Omega \pm c_1 k)^2 + \Gamma_1^2 a^2 k^4/4}\ ,
\label{eq:B.4}
\ee
with $\Gamma_1$ from Eq.~(\ref{eq:3.4c}).
The equal-time correlation function, which is, apart from a factor of $T$, equal to the $\psi_{1\pm}$-susceptibility $\chi_{\psi}$ by the 
fluctuation-dissipation theorem, is obtained by integrating over all frequencies:
\be
T\chi_{\psi} = \int_{-\infty}^{\infty} \frac{d\Omega}{2\pi}\ \langle \psi_{1\pm}\,\psi_{1\pm}^*\rangle_{\Omega,{\bm k}} = -2 T a (n_0/c_1)^2\ .
\label{eq:B.5}
\ee
Now consider $k \theta_1 = -i (c_1/2 n_0 a) (\psi_{1+} - \psi_{1-})$. Equation (\ref{eq:B.5}) yields $k^2 \chi_{\theta_1} = 1/a$, or
\be
\chi_{\theta_1} = 1/a k^2
\label{eq:B.6}
\ee
in agreement with Eq.~(\ref{eq:2.17b}). 

We see that the structure of the left eigenvector, which is very different from the right one, is crucial for obtaining the
correct result for the static susceptibility. The more complicated cases can be analyzed analogously. In particular, we
note that the structure of the left eigenvector ${\bm\nu}_{{\rm L}2,\pm}$ in the longitudinal-field case, Eq.~(\ref{eq:3.17g}), 
makes sure that there in no diverging static susceptibility, in agreement with the absence of any Goldstone modes in the
static analysis in Sec.~\ref{subsubsec:II.A.3}.


\begin{thebibliography}{45}
\bibitem[{FM_()}]{FM_footnote}
\bibinfo{note}{We do not refer to this case as ``coexisting AFM and
  ferromagnetic order" since the nonvanishing homogeneous magnetization can,
  and in general will, be due to ferrimagnetic rather than ferromagnetic
  order.}
\bibitem[{eig()}]{eigenvectors_footnote}
\bibinfo{note}{The degeneracy of the two eigenfrequencies at $k=0$ makes the
  determination of the eigenvectors nontrivial. The correct linear combinations
  of vectors that yield zero upon multiplication with the matrix, as shown in
  Eqs.~(\ref{eq:3.44b}, \ref{eq:3.44c}), have the property that the left and
  right eigenvectors belonging to different eigenfrequencies are orthogonal in
  the sense that ${\bm\nu}_{\rm{R}1,+}\cdot{\bm\nu}_{\rm{L}1,-} =
  {\bm\nu}_{\rm{R}1,-}\cdot{\bm\nu}_{\rm{L}1,+} = 0$. The analogous
  orthogonality property holds for the eigenvectors shown in
  Eqs.~(\ref{eq:3.17d}) - (\ref{eq:3.17g}); however, this case comes without
  the complications due to degenerate eigenfrequencies.}
\end{thebibliography}
\end{document}